\documentclass[12pt,english]{article}
\pdfoutput=1
\usepackage[T1]{fontenc}
\usepackage{graphicx,array}
\usepackage{cite}
\usepackage{color}
\usepackage{latexsym}
\usepackage{amsthm}
\usepackage{amsmath}
\usepackage[titletoc]{appendix}
\usepackage{enumitem}
\usepackage{amssymb}
\usepackage{color}
\usepackage[unicode=true,
 bookmarks=true,bookmarksnumbered=false,bookmarksopen=false,
 breaklinks=false,pdfborder={0 0 1},backref=false,colorlinks=true]
 {hyperref}
\hypersetup{pdftitle={Outer Entropy and Quasilocal Energy},
 pdfauthor={Raphael Bousso, Yasunori Nomura, Grant N. Remmen},
 citecolor=black,linkcolor=black,urlcolor=black}
\usepackage{breakurl}
\usepackage[hang,flushmargin]{footmisc} 

\def\simgt{\mathrel{\lower2.5pt\vbox{\lineskip=0pt\baselineskip=0pt
           \hbox{$>$}\hbox{$\sim$}}}}
\def\simlt{\mathrel{\lower2.5pt\vbox{\lineskip=0pt\baselineskip=0pt
           \hbox{$<$}\hbox{$\sim$}}}}

\newcommand{\be}{\begin{equation}}
\newcommand{\ee}{\end{equation}}
\newcommand{\bea}{\begin{eqnarray}}
\newcommand{\eea}{\end{eqnarray}}
\newcommand{\Eq}[1]{Eq.~(\ref{#1})}
\newcommand{\Eqs}[2]{Eqs.~(\ref{#1}) and (\ref{#2})}
\newcommand{\Sec}[1]{Sec.~\ref{#1}}

\newcommand{\Fig}[1]{Fig.~\ref{#1}}

\newcommand{\Ref}[1]{Ref.~\cite{#1}}

\newcommand{\So}[1]{S^{({\rm outer})}[#1]}

\newcommand*\oline[1]{%
  \vbox{%
    \hrule height 0.5pt
    \kern0.68ex
    \hbox{%
      \kern-0.1em
      \ifmmode#1\else\ensuremath{#1}\fi
      \kern-0.1em
    }
  }
}

\definecolor{nicered}{rgb}{0.7,0.1,0.1}
\definecolor{nicegreen}{rgb}{0.1,0.5,0.1}
\hypersetup{
 colorlinks,citecolor=black,linkcolor=black,urlcolor=black}
\usepackage{xcolor}

\begin{document}

\interfootnotelinepenalty=10000
\baselineskip=18pt
\hfill

\vspace{1.5cm}
\thispagestyle{empty}
\begin{center}
{\LARGE \bf
Outer Entropy and Quasilocal Energy
}\\
\bigskip\vspace{5mm}{
{\large Raphael Bousso,${}^a$ Yasunori Nomura,${}^{a,b}$ and Grant N. Remmen${}^a$}
} \\[5mm]
{\it ${}^a$Berkeley Center for Theoretical Physics, Department of Physics \\
  and Theoretical Physics Group, Lawrence Berkeley National Laboratory,\\
  University of California, Berkeley, CA 94720, USA \\[1.5mm]
${}^b$Kavli Institute for the Physics and Mathematics of the Universe (WPI),\\
 UTIAS, The University of Tokyo, Kashiwa, Chiba 277-8583, Japan
}
\let\thefootnote\relax\footnote{e-mail: 
\url{bousso@lbl.gov}, \url{ynomura@berkeley.edu}, \url{grant.remmen@berkeley.edu}}
\end{center}

\bigskip
\centerline{\large\bf Abstract}
\begin{quote} \small
We define the coarse-grained entropy of a ``normal'' surface $\sigma$, i.e., a surface that is neither trapped nor antitrapped. 
Following Engelhardt and Wall, the entropy is defined in terms of the area of an auxiliary extremal surface. 
This area is maximized over all auxiliary geometries that can be constructed in the interior of $\sigma$, while holding fixed the spatial exterior (the outer wedge). 
We argue that the area is maximized when the stress tensor in the auxiliary geometry vanishes, and we develop a formalism for computing it under this assumption. 
The coarse-grained entropy can be interpreted as a quasilocal energy of $\sigma$. This energy possesses desirable properties such as positivity and monotonicity, which derive directly from its information-theoretic definition.
\end{quote}
	
\setcounter{footnote}{0}

\newpage
\tableofcontents
\newpage

\section{Introduction}
\label{sec:Introduction}

The idea of coarse-graining---of integrating out microscopic degrees of freedom from an effective description of a system---is fundamental to thermodynamics.
The link between thermodynamics and geometry has been a crucial observation in the quest to understand quantum gravity since the discovery of Hawking radiation and the Bekenstein-Hawking entropy \cite{Hawking:1971tu,Bardeen:1973gs,Bekenstein:1972tm,Bekenstein:1973ur,Hawking:1974rv,Hawking:1974sw}.
The development of the holographic principle \cite{tHooft:1993dmi,Susskind:1994vu,Bekenstein:1980jp,Bousso:1999xy,Bousso:1999cb} and the AdS/CFT correspondence \cite{Maldacena:1997re,Gubser:1998bc,Witten:1998qj,Aharony:1999ti} has led to further insights into the geometric nature of gravitational entropy, including the Ryu-Takayanagi (RT) formula \cite{Ryu:2006bv,Ryu:2006ef,Lewkowycz:2013nqa} and its extension by Hubeny, Rangamani, and Takayanagi (HRT) \cite{Hubeny:2007xt,Wall:2012uf,Dong:2016hjy}, as well as various entropy bounds\cite{Bousso:1999xy,Bousso:1999cb,Casini:2008cr,Bousso:2014sda,Bousso:2014uxa}.
Nonetheless, an association of a calculable, coarse-grained entropic quantity with {\it arbitrary} surfaces has proved elusive.
In this paper, we make progress towards this goal, defining and calculating a coarse-grained holographic entropy for a large class of surfaces.

A recent proposal by Engelhardt and Wall (EW) \cite{Engelhardt:2017aux} clarifies the coarse-graining associated with the entropy of a black hole. If a black hole is formed from a pure state and we assume unitary evolution, then the fine-grained entropy vanishes. To associate an entropy to the area of the black hole, some form of coarse-graining is required.
The EW proposal applies not to the event horizon, but to any leaf $\sigma$ of a spacelike holographic screen. That is, $\sigma$ is marginally trapped (or antitrapped), and a locally spacelike hypersurface is foliated by a family of surfaces that includes $\sigma$ \cite{Bousso:2015mqa,Bousso:2015qqa}. Such a leaf can be thought of as a black hole boundary. Unlike the event horizon, its defining properties can be established from local data near $\sigma$.

EW propose to coarse-grain by holding fixed the exterior geometry of $\sigma$ but allowing an arbitrary geometry in the interior. One can then maximize the fine-grained entropy of this new spacetime to define an ``outer entropy.'' This can be made precise in the case where the exterior is asymptotic to anti-de Sitter spacetime. In this case the entropy is a von~Neumann entropy of the full quantum gravity theory, the boundary conformal field theory. It can be determined to leading order from the bulk geometry as the area of any stationary surface of minimal area that is homologous to the boundary.
Remarkably, the EW prescription naturally extends beyond the context of AdS/CFT: we can think of the coarse-grained entropy of any marginally-trapped surface $\sigma$ as the largest area of any minimal-area stationary surface that can be constructed when we allow the interior of $\sigma$ to vary.

In this paper, we will exploit another natural generalization of the EW proposal. One can vary the geometry and search for stationary surfaces inside of {\it any} surface $\sigma$, whether or not $\sigma$ is marginally trapped. To have a good notion of ``inside,'' we would like $\sigma$ to not be strictly trapped or antitrapped, but it need not be marginally trapped. The remaining possibility is simply that $\sigma$ is ``normal,'' i.e., that one of the orthogonal future-directed null congruences has everywhere positive expansion and the other one has everywhere negative expansion. In this case, the inside direction is the spacelike region on the negative-expansion side (see \Fig{fig:Cauchy}). Nomura and Remmen (NR)~\cite{Nomura:2018aus} previously formulated this generalization to normal surfaces in the case of spherically-symmetric spacetimes, but in this work we will consider general normal surfaces without assuming spherical symmetry.

An example of a normal surface is a sphere in empty Minkowski space. In fact, in this case the exterior region would be empty and the Arnowitt-Deser-Misner (ADM) mass~\cite{Arnowitt:1959ah} would vanish. Positive global mass~\cite{Schoen:1979rg,Witten:1981mf} then guarantees that the interior is vacuum Minkowski, and there cannot be another geometry with a nonzero stationary surface.  Another simple example is a round sphere outside of a Schwarzschild black hole. In this case the interior that maximizes the coarse-grained entropy is the maximally extended (``two-sided'') Schwarzschild solution of the same mass. The relevant stationary surface is the bifurcation surface of this solution.

From these examples, we can glean some key properties of the generalized construction that we will explore in this work. First, the coarse-grained entropy associated with a normal surface will {\em not} be equal to its area, but will be smaller. Physically, this makes sense, as a normal surface is normal because gravity is weaker. It does not enclose as much mass as a marginally-trapped surface of the same area. The largest black hole that can sit behind such a surface cannot be as large as the surface itself.

Since our construction will apply to normal surfaces, it includes the case of dynamical event horizons.
That is, we will be associating a coarse-grained entropy to the event horizon, though this entropy will not equal the horizon area.
This observation allows our construction to evade the no-go result of \Ref{Engelhardt:2017wgc}.

We will give an explicit geometric construction that identifies the stationary surface. Our construction can be thought of as finding the biggest two-sided black hole that might sit inside $\sigma$, if only the exterior is held fixed. This naturally leads to a quasilocal definition of energy associated with a normal surface $\sigma$, as an appropriate monotonic function of the area of the bifurcation surface of that black hole.

In the context of asymptotically AdS spacetimes, the generalized EW prescription is still a genuine coarse-graining, and we again expect this to generalize to other spacetimes. We will argue, though not prove, that our geometric construction succeeds in finding the interior geometry with the largest possible stationary surface, for a large class of surfaces $\sigma$. Then, as we consider a sequence $\sigma(r)$ of nested normal surfaces in the same geometry, the associated areas {\em must} be monotonic, simply because we hold less exterior data fixed as we move out to larger surfaces. The coarse-grained entropy, and hence the area, cannot decrease under such an operation. This establishes an important property that one would like a quasilocal energy to obey. Interestingly, the property does not hold for any obvious geometric reason at the level of the details of the algorithm, but is established here based on an information-theoretic argument.

This paper is organized as follows. 
In \Sec{sec:outer}, we review the motivation and definition of the outer entropy as a useful coarse-grained holographic quantity.
After discussing the characteristic initial data formalism, in \Sec{sec:construction} we give our procedure for constructing an HRT surface interior to a normal codimension-two surface.
We conjecture that this algorithm is optimal and therefore computes the outer entropy, and we present evidence for this conjecture in \Sec{sec:optimization}.
In \Sec{sec:quasilocal}, we use the outer entropy to define a quasilocal energy quantity and explore its relationship with other definitions of energy in general relativity.
Finally, in \Sec{sec:BTZ}, we consider the example of a codimension-two surface near which the geometry is locally that of the Ba\~{n}ados-Teitelboim-Zanelli (BTZ) metric~\cite{Banados:1992wn}, which will provide an illustrative example of our algorithm for a spacetime with rotation that nonetheless can be treated analytically.
We conclude with a discussion of future directions in \Sec{sec:discussion}.

\section{Outer Entropy}
\label{sec:outer}

Before presenting our construction of the maximal HRT surface, let us first carefully define our coarse-grained entropy and identify our assumptions.
Consider a quantum state defined on the disjoint union of a collection of closed spacelike manifolds 
having a classical bulk holographic dual spacetime obeying the Einstein 
equations.  The von~Neumann entropy $S[\rho] = -{\rm tr}\, \rho \log \rho$
associated with the reduced density matrix $\rho$ of some region $\Gamma$ 
is then given for the static case by the area of the RT 
surface and for general 
time-dependent spacetimes by that of the HRT 
surface:
\be
  S[\rho] = \frac{A[\text{HRT surface}]}{4G\hbar}.
\label{eq:HRTentropy}
\ee
The RT surface is simply the minimal-area surface on the relevant bulk spatial 
slice anchored to the boundary of $\Gamma$, while the HRT surface can be found 
using the maximin prescription of \Ref{Wall:2012uf}.  If the boundary state is 
pure, the entropy in \Eq{eq:HRTentropy} characterizes the entanglement between 
the subregion $\Gamma$ and the rest of the boundary state.  A case of particular 
interest is the entropy associated with an entire boundary manifold for 
a spacetime containing a wormhole.  In this case, the HRT surface $X_{\rm HRT}$ 
is homologous to the entire boundary region and has area characterizing the width 
of the wormhole throat.  Specifically, $X_{\rm HRT}$ is given by the closed, boundaryless, 
codimension-two surface for which the orthogonal null congruences have vanishing 
expansion and that has the area equal to the minimal cross section of some 
Cauchy slice.

A deeper understanding of coarse-graining and renormalization group flow is 
crucial to furthering our knowledge of holography, both within the AdS/CFT 
correspondence~\cite{Akhmedov:1998vf,Alvarez:1998wr,Balasubramanian:1999jd,%
Skenderis:1999mm,deBoer:1999tgo} and in the quest to generalize it to other 
spacetimes~\cite{Bousso:1999cb,Nomura:2016ikr,Nomura:2017npr,Nomura:2017fyh,%
Nomura:2018kji,Nomura:2018aus}.  A quantity of particular interest is the 
{\it outer entropy}~\cite{Engelhardt:2017aux,Nomura:2018aus} associated with 
a codimension-two surface $\sigma$:
\be
  \So{\sigma} = \max_{\tilde\rho} 
    \left(S[\tilde\rho]\,:\,O_W(\sigma)\,\text{fixed}\right),
\label{eq:outerentropydef}
\ee
where $O_W(\sigma)$ is the {\it outer wedge}, the subset of the spacetime in 
the interior of the domain of dependence of the partial Cauchy surface connecting 
$\sigma$ with the boundary.  The maximization in \Eq{eq:outerentropydef} is 
computed over CFT states $\tilde \rho$ defined on the outer boundary of $O_W(\sigma)$ for 
which the geometry in $O_W(\sigma)$ is fixed.  In the case of a pure state defined on two disconnected 
boundaries, the outer entropy of one of the boundaries computes its maximum 
entanglement entropy with the other boundary, subject to the constraint that 
the relevant outer wedge have fixed geometry.

In geometric terms, the outer entropy is given by ($1/4G\hbar$ times) the area 
of the largest HRT surface one can put inside%
\footnote{One can show that, if it is possible to construct an HRT surface in 
 a geometry while keeping $O_W(\sigma)$ fixed, with $\sigma$ being a normal or 
 marginally-trapped surface homologous to the boundary and for which a partial Cauchy surface exists connecting $\sigma$ with the boundary such that any slice subtending $\sigma$ has greater area than $\sigma$, then the HRT surface is in (the closure of) the 
 domain of dependence of the interior of $\sigma$~\cite{Nomura:2018aus}.
 \label{locationHRT}}
the surface $\sigma$, given its fixed exterior geometry.  The outer entropy is 
a coarse-grained quantity in holography; we have in effect coarse-grained over 
all information about the spacetime except for the geometry on $O_W(\sigma)$. 
Note that we do not need the full apparatus of AdS/CFT for this coarse-grained 
interpretation of the outer entropy.  We only need the assumptions of 
Refs.~\cite{Engelhardt:2017aux,Nomura:2018aus} that the HRT surface constitutes 
a fine-grained (i.e., von~Neumann) entropy associated with the reduced density matrix 
in the relevant region on the boundary.

EW argued that if $\sigma$ is a marginally-trapped or -antitrapped surface, then $\So{\sigma} = A[\sigma]/4G\hbar$.  Given 
the area law for holographic screens~\cite{Bousso:2015mqa,Bousso:2015qqa}, this 
implies a thermodynamic second law associated with the evolution of the entropy 
along the holographic screen.  NR~\cite{Nomura:2018aus} generalized 
the concept of a holographic screen to a particular class of surfaces that are 
not marginally trapped or antitrapped, including the event horizon.  It was 
shown there that these generalized holographic screens also satisfy an area 
law and, for spherically-symmetric surfaces, a second law for the outer entropy 
(despite the fact that $\So{\sigma} \neq A[\sigma]/4G\hbar$ for surfaces that 
are not marginally trapped or antitrapped).  For a normal surface, one can 
show~\cite{Nomura:2018aus,Engelhardt:2018kcs} using the Raychaudhuri equation 
that the outer entropy is upper bounded by the area
\be
  \So{\sigma} < \frac{A[\sigma]}{4G\hbar}.
\ee
In the following sections, we will compute the outer entropy for a normal 
surface $\sigma$, subject to certain assumptions, providing an algorithm for 
computing this coarse-grained holographic quantity in generality.  Unlike in 
EW~\cite{Engelhardt:2017aux}, $\sigma$ need not be marginally trapped or antitrapped, 
and unlike in NR~\cite{Nomura:2018aus}, we will not assume spherical symmetry. 
Later, we will argue that the outer entropy can be viewed as a compelling 
quasilocal energy in general relativity.

\section{Construction of the Spacetime}
\label{sec:construction}

Having noted the general upper bound for $\So{\sigma}$, we will seek a lower 
bound on the outer entropy by explicitly constructing a spacetime consistent 
with $O_W(\sigma)$ and computing the area of the HRT surface $X_{\rm HRT}$ 
in this spacetime.  Later, we will argue that the choices we make in this 
construction maximize $A[X_{\rm HRT}]$, so that this ``lower bound'' 
actually equals $\So{\sigma}$ itself.  The general approach to the construction, 
as well as our notation, will closely follow that of NR~\cite{Nomura:2018aus}.  However, because of important differences that occur in the 
nonspherical case as well as for self-consistency, we will review the formalism 
here before presenting the details of the construction.

\subsection{Characteristic initial data formalism}
\label{subsec:CIDF}

\begin{figure}[t]
\begin{center}
\hspace{-5mm} \includegraphics[width=8cm]{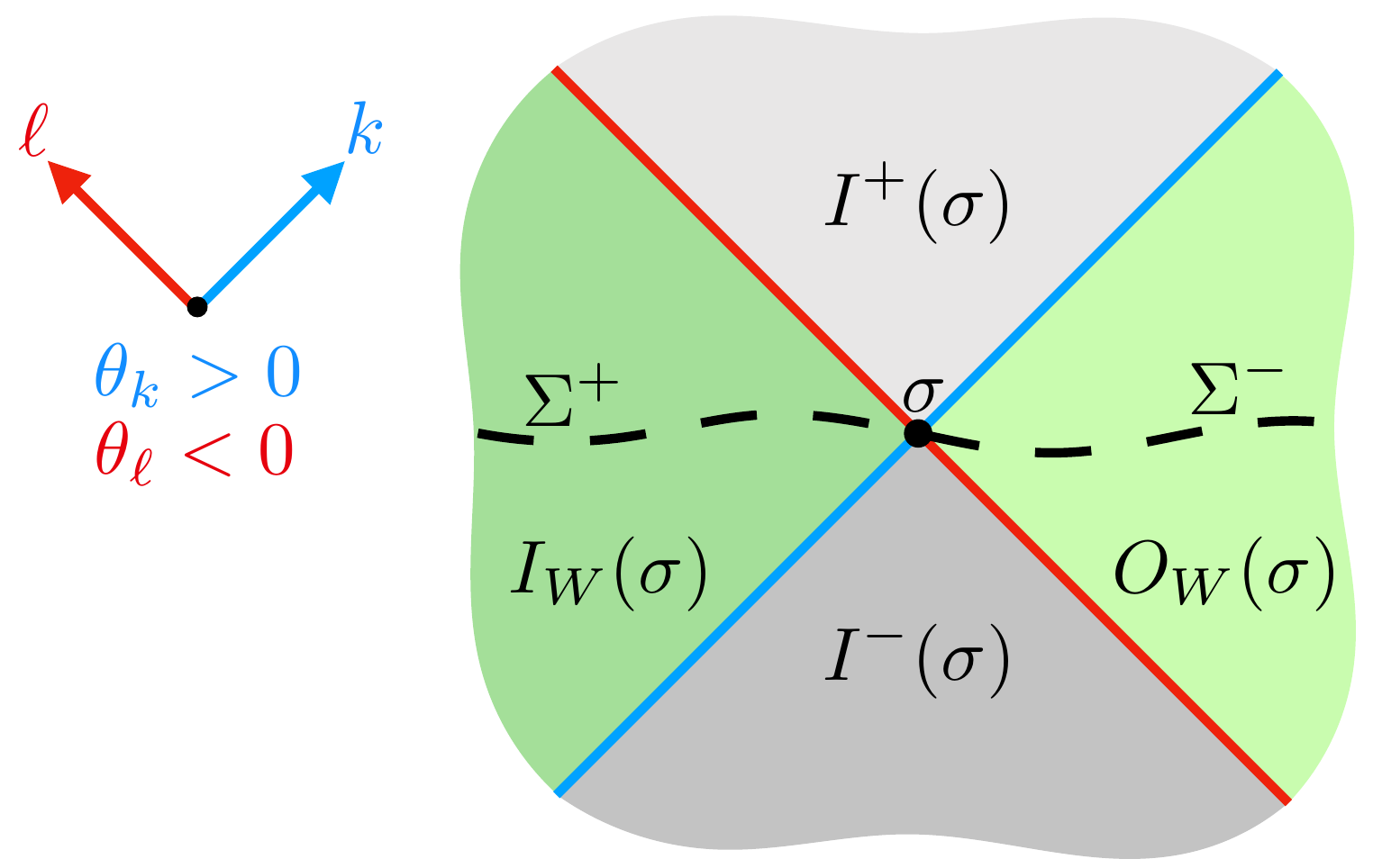}
\end{center}
\vspace{-5mm}
\caption{Penrose diagram illustrating a normal codimension-two surface 
$\sigma$, with $\theta_k > 0$ and $\theta_\ell < 0$, that splits a Cauchy 
surface into an inner ($\Sigma^+$) and outer ($\Sigma^-$) portion.  The 
light sheets $N_k(\sigma)$ (blue) and $N_\ell(\sigma)$ (red) defined in 
\Eq{eq:lightsheetdef} split the spacetime into four pieces~\cite{Akers:2017nrr}: 
the past and future $I^\pm(\sigma)$, the inner wedge $I_W(\sigma) = 
\mathring D(\Sigma^+(\sigma))$ and the outer wedge $O_W(\sigma) = 
\mathring D(\Sigma^-(\sigma))$.}
\label{fig:Cauchy}
\end{figure}

Let us first review some notation and geometrical formalism.  Throughout, any 
spacetime $({\cal M},g_{ab})$ that we consider will be taken to be globally 
hyperbolic, supplemented with appropriate boundary conditions for spacetimes 
with boundary~\cite{Avis:1977yn}.  Given our codimension-two, compact, boundaryless, acausal 
surface $\sigma$, there are two future-directed orthogonal null congruences 
with tangent vectors that we label $k$ and $\ell$.  We can arbitrarily label 
$k$ to be the ``outgoing'' congruence and $\ell$ the ``ingoing'' congruence, 
and for any Cauchy surface $\Sigma$ split by $\sigma$ into two pieces 
$\Sigma^\pm$ with $\sigma = \dot{\Sigma}^+ = \dot{\Sigma}^-$, we take 
$\Sigma^-$ (the exterior) to lie in the direction of $k$ and $\Sigma^+$ 
(the interior) to lie in the direction of $\ell$.%
\footnote{We choose this notation for consistency with 
 Refs.~\cite{Bousso:2015qqa,Nomura:2018aus}.  Throughout, we 
 use the standard notation of $I^\pm$ for the chronological future 
 and past, $D^\pm$ for the future and past domains of dependence, 
 $D(S) = D^+(S)\cup D^-(S)$, and $\dot S$, $\mathring S$, and $\overline S$ 
 for the boundary, interior, and closure of a set $S$, respectively. Our notation for arguments is as follows: square brackets for a quantity defined as a functional of some subset of points in ${\cal M}$ (e.g., $A[\sigma]$), round brackets for arguments on objects that are themselves subsets of ${\cal M}$ (e.g., $D(S)$), and round brackets for scalar arguments in functions.}
In this notation, the outer wedge is $O_W(\sigma) = \mathring 
D(\Sigma^-(\sigma))$.  We define the light sheets originating 
from $\sigma$ as in Refs.~\cite{Nomura:2018aus,Bousso:2015qqa}:
\be
\begin{aligned}
  N_{+k}(\sigma) &= \dot{I}^+(\Sigma^+) - \Sigma^+ 
    = \dot{D}^+(\Sigma^-) - I^-(D^+(\Sigma^-)) \\
  N_{-k}(\sigma) &= \dot{I}^-(\Sigma^-) - \Sigma^- 
    = \dot{D}^-(\Sigma^+) - I^+(D^-(\Sigma^+)) \\
  N_{+\ell}(\sigma) &= \dot{I}^+(\Sigma^-) - \Sigma^- 
    = \dot{D}^+(\Sigma^+) - I^-(D^+(\Sigma^+)) \\
  N_{-\ell}(\sigma) &= \dot{I}^-(\Sigma^+) - \Sigma^+ 
    = \dot{D}^-(\Sigma^-) - I^+(D^-(\Sigma^-))
\end{aligned}
\label{eq:lightsheetdef}
\ee
and define $N_k(\sigma) = N_{+k}(\sigma)\cup N_{-k}(\sigma)$ and similarly 
$N_\ell(\sigma) = N_{+\ell}(\sigma)\cup N_{-\ell}(\sigma)$.  See \Fig{fig:Cauchy} 
for a summary of the definitions for how $\sigma$ splits the spacetime.  The 
vector $k$ is parallel transported along $N_k(\sigma)$ and, similarly, $\ell$ 
is parallel transported along $N_\ell(\sigma)$.  Along $N_k(\sigma)$, $\ell$ 
is parallel transported but continually rescaled such that $k\cdot \ell = -1$, 
and $k$ is similarly defined on $N_\ell$.  Having made these choices, we can 
define null vector fields everywhere in ${\cal M}$ such that $k$ and $\ell$ 
are each parallel transported along themselves and $k\cdot \ell = -1$.

The induced metric on $\sigma$ is
\be
  q_{ab} = g_{ab} + 2k_{(a} \ell_{b)},
\ee
where throughout we use the normalized convention for (anti-)symmetrization, 
$T_{(ab)} = \frac{1}{2}(T_{ab} + T_{ba})$.  Using the induced metric as 
a projector (where we raise indices on $q_{ab}$ using the full metric 
$g_{ab}$), we can define the null extrinsic curvature in the standard 
manner~\cite{Wald,Bousso:2016fia},
\be
\begin{aligned}
  (B_k)_{ab}    &= q_a^{\;\;c} q_b^{\;\;d} \nabla_d k_c   \\
  (B_\ell)_{ab} &= q_a^{\;\;c} q_b^{\;\;d} \nabla_d \ell_c,
\end{aligned}
\ee
from which we can define the null expansions
\be
\begin{aligned}
  \theta_k    &= q^{ab} (B_k)_{ab}   \\
  \theta_\ell &= q^{ab} (B_\ell)_{ab}
\end{aligned}
\ee
and the shears
\be
\begin{aligned}
  (\varsigma_k)_{ab}    &= (B_k)_{(ab)} - \frac{1}{D-2} \theta_k q_{ab}      \\
  (\varsigma_\ell)_{ab} &= (B_\ell)_{(ab)} - \frac{1}{D-2} \theta_\ell q_{ab},
\end{aligned} 
\ee
where $D$ is the dimension of the spacetime.  Since we are considering 
hypersurface orthogonal geodesics, $B_k$ and $B_\ell$ are symmetric tensors. 
We choose $\sigma$ to be a normal surface, i.e., one on which $\theta_k > 0$ 
and $\theta_\ell < 0$.  For spacetimes with boundary, we will further require 
that $\sigma$ be chosen to be homologous to the boundary and such that there 
exists a Cauchy surface $\Sigma$ for which every slice of $\Sigma^-$ subtending 
$\sigma$ has area larger than that of $\sigma$.

Given a Cauchy surface formed by a collection of null surfaces, 
the characteristic initial data formalism \cite{Rendall:2000,%
Brady:1995na,ChoquetBruhat:2010ih,Luk:2011vf,Chrusciel:2012ap,%
Chrusciel:2012xf,Chrusciel:2014lha} guarantees that one can uniquely 
specify a spacetime from data on the Cauchy surface alone, provided that the data 
satisfy a set of constraint equations.  In particular, for null surfaces 
formed by $N_k(\sigma)$ for some surface $\sigma$, the constraint equations 
are~\cite{Hayward:1993wb,Hayward:2004fz,Hayward:2006ss,Gourgoulhon:2005ng,%
Cao:2010vj,Price:1986yy}
\be
\begin{aligned}
  \nabla_k \theta_k &= -\frac{1}{D-2} \theta_k^2 
    - \varsigma_k^2 - 8 \pi G\,T_{kk} \\
  q_a^{\;\;b} {\cal L}_k \omega_b &= -\theta_k \omega_a 
    + \frac{D-3}{D-2}{\cal D}_a \theta_k - ({\cal D}\cdot \varsigma_k)_a 
    + 8 \pi G\,T_{ak} \\
  \nabla_k \theta_\ell &= -\frac{1}{2}{\cal R} - \theta_k \theta_\ell 
    + \omega^2 + {\cal D}\cdot \omega + 8 \pi G\,T_{k\ell} + \Lambda.
\end{aligned}
\label{eq:constraintk}
\ee
For $N_\ell(\sigma)$, the constraint equations are the same as in 
\Eq{eq:constraintk}, but with $k \leftrightarrow \ell$ and $\omega 
\rightarrow -\omega$.  Here, $\omega_a$ is the twist one-form gauge 
field defined as \cite{Hayward:2006ss,Bousso:2016fia}
\be
  \omega_a = \frac{1}{2} q_{ab} {\cal L}_k \ell^b 
  = -\ell_b \,q_a^{\;\;c} \nabla_c k^b,
\label{eq:twistdef}
\ee
${\cal R}$ is the intrinsic Ricci curvature on slices of the congruence at 
constant affine parameter, ${\cal D}_a = q_a^{\;\;b}\nabla_b$ is the covariant 
derivative along $\sigma$, ${\cal L}_k$ denotes the Lie derivative along 
$k$, and $k$ and $\ell$ as index subscripts denote 
indices contracted into $k^a$ and $\ell^a$, respectively.  The expansion 
and twist are required to be continuous across junctions, but the shears 
are not~\cite{Engelhardt:2018kcs,Wald,Luk:2012hi,Luk:2013zr}.  In \Eq{eq:constraintk}, the first line is the Raychaudhuri equation, the second 
is the Damour-Navier-Stokes~(DNS) equation, and the third is the cross-focusing 
equation, where we have substituted in the Einstein equations,
\be
  R_{ab} - \frac{1}{2}R\,g_{ab} + \Lambda\,g_{ab} = 8\pi G\,T_{ab}.
\ee

\subsection{Building an HRT surface}
\label{subsec:buildHRT}

Let us use the formalism discussed in \Sec{subsec:CIDF} to construct a spacetime 
that contains both $O_W(\sigma)$ and an HRT surface.  We will 
want to calculate the area of this HRT surface.
For reasons that will become clear later, we will choose initial data in 
the interior of $\sigma$, specifically on $N_{-k}(\sigma) \cup D(\Sigma^+)$, 
to satisfy%
\footnote{As discussed in \Ref{Nomura:2018aus}, we can set $T_{kk}$ and 
 $T_{k\ell}$ to zero along $N_{-k}(\sigma)$ consistent with our energy 
 conditions and energy-momentum conservation via a limiting procedure, 
 and a similar argument applies for $T_{\ell\ell}$.  Moreover, we can set 
 $\varsigma_k$ and $\varsigma_\ell$ to zero discontinuously via a shock wave 
 in the Weyl tensor~\cite{Wald}, which has no effect on $T_{ab}$.  As we will 
 see in \Sec{sec:optimization}, a consequence of the $\Lambda$-subtracted 
 dominant energy condition is that requiring $T_{kk} = T_{k\ell} = 
 T_{\ell\ell} = 0$ implies that $T_{ab} = 0$ in all components; see 
 footnote~\ref{Tabzero}.}
\be 
  T_{kk} = T_{\ell\ell} = T_{k\ell} = \varsigma_k = \varsigma_\ell = 0.
\label{eq:zeros}
\ee
The constraint equations \eqref{eq:constraintk} along $N_k(\sigma)$ then become\footnote{As shown in \Ref{Gourgoulhon:2005ng}, $q_a^{\;\;b}\nabla_k \omega_b 
= q_a^{\;\;b} {\cal L}_k \omega_b - (B_k)_a^{\;\;b} \omega_b$.  By definition, 
$\nabla_k \omega_a = \partial_k \omega_a - \Gamma^b_{ak} \omega_b$, where 
$\Gamma^a_{bc}$ are the Christoffel symbols.  Since we are contracting 
$\Gamma^b_{ak}$ with $\omega_b$ and ultimately projecting the lower index 
using $q$, we are interested in $\Gamma_{ak}^b$ where both $a$ and $b$ point 
along $\sigma$.  Since $g_{ak}=0$ identically for $a$ pointing along $\sigma$ 
(since $k^a$ is orthogonal to $\sigma$) and since $g_{kk}=0$ and $g_{k\ell}=-1$, 
we have $\Gamma^b_{ak} = \frac{1}{2}g^{bc}\partial_k g_{ac}$.  The partial 
$k$ derivative of the transverse components of the metric is dictated simply 
by the expansion $\theta_k$, so for $a$ and $b$ pointing along $\sigma$, 
$\partial_k g_{ab} = \frac{2}{D-2}\theta_k g_{ab}$ and hence $\Gamma^b_{ak} 
= \frac{1}{D-2}\theta_k \delta^b_a$. Thus, in our coordinate system, $
  q_a^{\;\;b} {\cal L}_k \omega_b = q_a^{\;\;b}\partial_k \omega_b 
    + (\varsigma_k \cdot \omega)_a$
and similarly for $q_a^{\;\;b} {\cal L}_\ell \omega_b$.
Since every term on the right-hand side of the DNS equation in \Eq{eq:constraintksimp} points along $\sigma$, we can drop the projector $q_a^{\;\;b}$ from the left-hand side.\label{longfootnote}}
\be
\begin{aligned}
  \partial_k \theta_k &= -\frac{1}{D-2}\theta_k^2 \\
  \partial_k \omega_a &= -\theta_k \omega_a +\frac{D-3}{D-2}{\cal D}_a \theta_k \\
  \partial_k \theta_\ell &= -\frac{1}{2}{\cal R} - \theta_k \theta_\ell 
    + \omega^2 + {\cal D}\cdot \omega + \Lambda.
\label{eq:constraintksimp}
\end{aligned}
\ee
Let us define an affine parameter $\nu$ on $N_k(\sigma)$, with $\nu = 0$ 
corresponding to $\sigma$ and normalized such that $k^a=({\rm d}/{\rm d}\nu)^a$. 
We will write the coordinates on $\sigma$ as $x^i$.  On constant-$\nu$ slices 
$Y(\nu)$ of $N_k(\sigma)$, we can define coordinates $x^i$ via the exponential 
map from $\sigma$.  Namely, the $x^i$ coordinates of a point $y \in X(\nu)$ 
are defined to be the coordinates of the point $z\in \sigma$ for which the 
orthogonal null geodesic in the $k$ direction originating from $z$ passes 
through $y$.%
\footnote{By the theorem of \Ref{Akers:2017nrr}, which characterizes 
 $N_k(\sigma)$, this map is bijective unless $y$ is at a caustic or 
 nonlocal intersection of null geodesics.}

We wish to construct a spacetime that has an extremal surface $X_{\rm HRT}$, 
for which both of the null congruences orthogonal to $X_{\rm HRT}$ vanish. 
First, we use the constraint equations to locate a surface $Y_0$ along 
$N_{-k}(\sigma)$ on which $\theta_\ell$ vanishes.  Note that, a priori, 
this condition does not make $Y_0$ a marginally-antitrapped surface: the 
ingoing null congruence orthogonal to $Y_0$ has tangent vector $\tilde \ell$, 
which is not in general the same as $\ell$, since the affine parameter 
$\nu_0(x^i)$ defining $Y_0$ can vary as a function of $x^i$, while $\ell$ 
is orthogonal to constant-$\nu$ slices of $N_k(\sigma)$.  There should, 
however, be some marginally-antitrapped surface $Y_{\rm MA}$ near $Y_0$, 
on which $\theta_{\tilde{\ell}}=0$.  The relation between $\theta_\ell [Y_0]$ 
and $\theta_{\tilde{\ell}}[Y_0]$ can be written as a second-order differential 
equation for $\nu_{0}(x^{i}$) (see, for example, \Ref{Engelhardt:2018kcs} 
for how this works in the special case of a light sheet with $\theta_k = 0$ 
everywhere).  One could then try to locate the surface $Y_{\rm MA}$ by solving 
this equation and optimize its area.

There is, however, a different way to address the problem.  Since the computation 
of the outer entropy can be performed under any gauge condition, we may choose 
a convenient gauge.  Specifically, we can {\it require} 
that $Y_0$ be a surface of constant affine parameter.  The gauge freedom 
allowing us to impose this condition is the $x^{i}$-dependent rescaling 
of $k^a$ on $\sigma$ (and concomitant inverse rescaling of $\ell^a$ so as 
to keep $k\cdot \ell = -1$).  With this condition, $\ell = \tilde{\ell}$ 
on $Y_0$, so that $Y_0$ is indeed a surface on which $\theta_{\ell} = 
\theta_{\tilde{\ell}}=0$; namely, $Y_0 = Y_{\rm MA}$ in this gauge.  Of 
course, we do not know a priori the proper gauge condition to guarantee 
this.  However, we can still find $Y_0$ under an arbitrary gauge choice, 
optimize the area of $Y_0$, and at the end select the gauge condition that 
makes $\nu_0$ constant.  Because of the optimization involved, this is 
equivalent to finding the optimal $Y_{\rm MA}$ using a prefixed gauge. 
This is the approach we will follow in the remainder of this section.

Once $Y_{\rm MA}$ is found, we can follow a null congruence toward the future 
along $N_{+\ell}(Y_{\rm MA})$.  Recalling our choices in \Eq{eq:zeros}, the 
constraint equations along $N_{+\ell}(Y_{\rm MA})$ are
\be
\begin{aligned}
  \partial_\ell \theta_\ell &= -\frac{1}{D-2}\theta_\ell^2 \\
  \partial_\ell \omega_a &= -\theta_\ell \omega_a 
    - \frac{D-3}{D-2}{\cal D}_a \theta_\ell \\
  \partial_\ell \theta_k &= -\frac{1}{2}{\cal R} - \theta_k \theta_\ell 
    + \omega^2 - {\cal D}\cdot \omega + \Lambda.
\label{eq:constraintlsimp}
\end{aligned}
\ee
On $N_{+\ell}(Y_{\rm MA})$, we choose to hold ${\cal R}$, $\omega_a$, and 
$\theta_\ell$ fixed along $\ell$ (the last of which vanishes).  The Raychaudhuri 
and DNS equations in \Eq{eq:constraintlsimp} are then trivially satisfied. 
Then, provided $\partial_\ell \theta_k[Y_{\rm MA}] < 0$, we eventually reach a 
surface $X_0$ on which $\theta_\ell = \theta_k = 0$.  Define $\Sigma_1 = 
N_{-k}(\sigma) \cap N_{+k}(Y_{\rm MA})$.  Moving along $\Sigma_1$ from 
$Y_{\rm MA}$ to $\sigma$, the area of cross sections strictly increases 
(since $\theta_k > 0$).  Thus, recalling that $\sigma$ is by definition 
a surface of minimal cross section on $\Sigma^-$, we find that $Y_{\rm MA}$ 
is a surface of minimal cross section on $\Sigma_1 \cup \Sigma^-$, so 
$Y_{\rm MA}$ satisfies the conditions of a ``minimar'' surface as defined 
in \Ref{Engelhardt:2018kcs}.

\begin{figure}[t]
\begin{center}
\hspace{-5mm} \includegraphics[width=12cm]{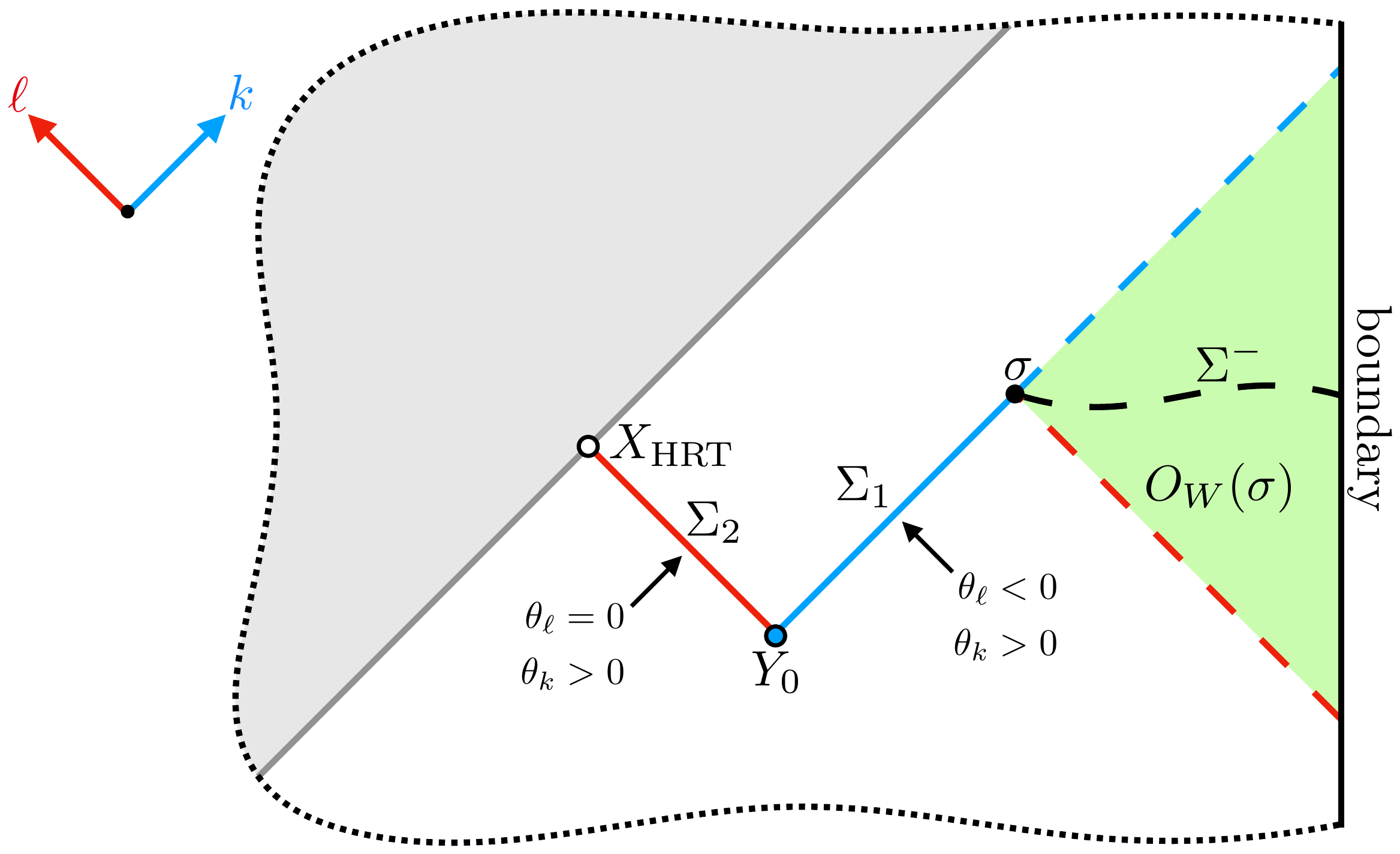}
\end{center}
\vspace{-5mm}
\caption{Portion of a Penrose diagram illustrating our construction of an HRT 
 surface realizing the outer entropy.  Holding the outer wedge (green) fixed, 
 we choose data on $N_{-k}(\sigma)$ (blue line) as described in text until 
 we reach a surface $Y_0$ (blue dot) on which $\theta_\ell = 0$.  We choose 
 a gauge such that $Y_0$ is marginally antitrapped, $Y_0 = Y_{\rm MA}$.  Again 
 choosing data as described in \Sec{subsec:buildHRT}, we follow the light 
 sheet $N_{+\ell}(Y_0)$ (red line) until we reach a surface $X_0$ on which  
 $\theta_k = 0$, provided $\partial_\ell \theta_k < 0$ on $Y_{\rm MA}$, which 
 we assume.  As discussed in the text, the existence of $X_0$ guarantees the 
 existence of an HRT surface $X_{\rm HRT}$ (white dot) on $N_{+k}(Y_0)$.  The 
 entire spacetime is completed (gray shading) by CPT-reflecting the initial 
 value data on $\Sigma^- \cup \Sigma_1 \cup \Sigma_2$.}
\label{fig:construction}
\end{figure}

Even though $\theta_k$ and $\theta_\ell$ vanish there, we cannot conclude 
that $X_0$ is an HRT surface.  Just as in the case of $Y_0$, the outgoing null 
geodesic congruence from $X_0$ has some tangent $\tilde k$, which may differ 
from $k$, so $\theta_{\tilde k}$ does not necessarily equal $\theta_k$. 
However, using the time-reverse of the construction in \Ref{Engelhardt:2018kcs}, 
the fact that $Y_0$ is a minimar surface guarantees that, along $N_{+\ell}(Y_0)$, 
there is some surface $X$ for which $\theta_{\tilde k}$ vanishes and for 
which $A[X]=A[X_0] = A[Y_{\rm MA}]$. (The details of how this construction 
works involve inverting the stability operator relating $\theta_k$ and 
$\theta_{\tilde k}$.)  To show that $X$ is indeed an HRT surface, it 
remains to exhibit a partial Cauchy surface homologous to the boundary 
on which $X$ is a minimal cross section.  Such a surface is $\Sigma_0^- 
= \Sigma^- \cup \Sigma_1 \cup \Sigma_2$, where $\Sigma_2$ is the portion of $N_{+\ell}(Y_{\rm MA})$ between $Y_{\rm MA}$ and $X$.  It follows that $X$ is a bona~fide HRT surface, 
with area equal to $A[X_0]$.%
\footnote{Let $\Sigma_0$, formed by $\Sigma_0^-$ and its CPT conjugate, 
 be the Cauchy surface for a spacetime that one constructs using the 
 characteristic initial data formalism.  For any extremal surface $\hat X$ 
 in this spacetime, with orthogonal null congruences with tangents $\hat k$ 
 and $\hat \ell$, one would find by the Raychaudhuri equation and the null 
 energy condition (NEC) that slices of $N_{\hat \ell}(\hat X)$ have area at 
 most $A[\hat X]$.  Hence, $A[\hat X] \geq A[X]$.}
We will denote this fact by writing $X$ as $X_{\rm HRT}$ henceforth.  See 
\Fig{fig:construction} for an illustration of our construction.

To find the expression for $A[X_{\rm HRT}]$, we still need to construct 
the appropriate surface $Y_0$ by solving the constraint equations on 
$N_{-k}(\sigma)$.  We now turn to this problem.

\subsection{Solution to the constraint equations}
\label{subsec:solution}

Let us solve the constraint equations \eqref{eq:constraintksimp}, given our 
choice \eqref{eq:zeros} of initial data.  By inverting the Raychaudhuri equation, 
we can solve $\theta_k(\nu)$ at $x^i$ as a function of $\theta_k[\sigma]$ at 
the same $x^i$:
\be
  \theta_k(\nu) = \left[ \frac{1}{\theta_k[\sigma]}+\frac{\nu}{D-2} \right]^{-1}.
\label{eq:thetaksol}
\ee
We will leave the $x^i$ arguments implicit everywhere.  We find it convenient 
to introduce a new variable $\xi$, a function of $\nu$ and $x^i$, to 
parameterize distance along $N_k(\sigma)$, defined by
\be
  \xi(\nu) = \frac{\theta_k(\nu)}{\theta_k[\sigma]} 
  = \left[ 1+\frac{\nu \theta_k[\sigma]}{D-2}\right]^{-1}.
\ee
We note that $\xi = 1$ corresponds to $\sigma$ and $\xi > 1$ corresponds to 
slices of $N_{-k}(\sigma)$. In terms of $\xi$, the derivative operator is
\be
  \partial_k = \frac{\partial \xi}{\partial\nu}\partial_\xi 
  = \frac{\partial_k \theta_k}{\theta_k[\sigma]}\partial_\xi 
  = -\frac{1}{D-2}\xi^2 \theta_k[\sigma]\partial_\xi.
\ee

We can write $\theta_k$ in the Raychaudhuri equation in \Eq{eq:constraintksimp} 
as $\xi\theta_k [\sigma]$ and, since
\be
  {\cal D}_i\xi 
  = -\frac{\frac{\nu}{D-2}}{\left[1+\frac{\nu\theta_k[\sigma]}{D-2}\right]^{2}} 
    {\cal D}_i \theta_k[\sigma] 
  = \left(\xi^2-\xi\right) {\cal D}_i \log\theta_k[\sigma],
\label{eq:Dxi}
\ee
we have the nice expression
\be
  {\cal D}_i\theta_k(\xi) = \xi {\cal D}_i \theta_k[\sigma] 
    + \theta_k[\sigma] {\cal D}_i \xi 
  = \xi^2 {\cal D}_i \theta_k[\sigma].
\ee
Hence, the DNS equation in \Eq{eq:constraintksimp} becomes
\be
  \xi\partial_\xi \omega_i 
  = (D-2) \omega_i - (D-3) \xi {\cal D}_i \log\theta_k[\sigma],
\ee
which has solution
\be
  \omega_i(\xi) = \omega_i[\sigma] \xi^{D-2} 
    + \left(\xi-\xi^{D-2}\right) {\cal D}_i \log\theta_k[\sigma].
\label{eq:omegasol}
\ee
By \Eq{eq:thetaksol} we have satisfied the Raychaudhuri equation in 
\Eq{eq:constraintksimp}, and by \Eq{eq:omegasol} we have satisfied 
the DNS equation.  It remains to compute the terms in the cross-focusing 
equation to solve for $\theta_\ell$ as a function of $\xi$.  Let us 
consider each term in turn.

Since $\partial_k g_{ij} = \frac{2}{D-2}\theta_k g_{ij}$, we 
have $\partial_k \log{\cal R} = -\frac{2}{D-2}\theta_k$, or equivalently, 
$\xi \partial_\xi\log{\cal R} 
= 2$, so
\be
  {\cal R}(\xi) = \xi^2 {\cal R}[\sigma].
\ee
Similarly, $\partial_{k}g^{ij} 
= -\frac{2}{D-2}\theta_k g^{ij}$ as shown in footnote~\ref{longfootnote}, so $\xi\partial_\xi g^{ij} = 2g^{ij}$, 
which has solution $g^{ij}(\xi) = \xi^2 g^{ij}[\sigma]$.  (Here, $i$ and $j$ 
are transverse indices, so we could write $q_{ij}$ everywhere for $g_{ij}$ 
in this statement.)  Since $\omega^i(\xi) = g^{ij}(\xi) \omega_j(\xi)$, we therefore obtain
\be
\begin{aligned}
  \omega^2(\xi) &= \xi^2 g^{ij}[\sigma] \omega_i(\xi) \omega_j(\xi) \\
  &= \xi^{2(D-1)} \omega^2[\sigma] + 2\left(\xi^3-\xi^D\right) 
    \xi^{D-2} \omega^i[\sigma] {\cal D}_{i}\log\theta_k[\sigma]   \\
    &\qquad + \left(\xi^2-\xi^{D-1}\right)^2 {\cal D}^i \log\theta_k[\sigma] 
    {\cal D}_i \log\theta_k[\sigma].
\end{aligned}
\ee
We can similarly compute ${\cal D}\cdot\omega$ as a function of $\xi$.\footnote{
For a one-form $v_a$ pointing along $\sigma$, $\nabla_a v_b = \partial_a v_b 
- \Gamma_{ba}^c v_c$, so $q^{ab} \nabla_a v_b$ contains only the transverse 
Christoffel symbols $\Gamma_{ab}^c$, where $a, b, c$ point along $\sigma$. 
But $\Gamma_{ab}^c(\nu) = \Gamma_{ab}^c[\sigma]$ since $g_{ki} = g_{\ell i} = g^{ki} = g^{\ell i}
= 0$ for transverse index $i$.  Hence, $(\nabla_a v_b)(\nu)$ is simply 
$\nabla_a(v_b(\nu))$ and so ${\cal D} \cdot v = q^{ab}\nabla_a v_b$ changes 
only as a result of the $\nu$-dependence of $v_a$.}  Recalling the expression 
in \Eq{eq:Dxi} for ${\cal D}_{i}\xi$ and the fact that $q^{ab}(\xi) = \xi^2 
q^{ab}[\sigma]$, we have
\be
\begin{aligned}
  {\cal D}\cdot\omega(\xi) &= \left(\xi^2-\xi\right)\left[\xi^2 - (D-2)\xi^{D-1}\right] 
    \left({\cal D}_i \log\theta_k [\sigma]\right)^2 
 \\&\qquad  + (D-2)\left(\xi^{D+1}-\xi^D\right) \omega^i[\sigma]{\cal D}_i \log\theta_k [\sigma] \\
 &\qquad + \left(\xi^3-\xi^D\right) \Box\log\theta_k[\sigma] 
    + \xi^D {\cal D}\cdot\omega[\sigma],
\end{aligned}
\ee
where $\Box={\cal D}\cdot{\cal D}$.

Let us define $q(\nu,x^{i})$ such that
\be
  \theta_\ell = \theta_\ell[\sigma] \frac{q}{\xi}.
\ee
Then
\be 
  \partial_k \theta_\ell 
  = -\frac{1}{D-2} \xi^2 \theta_k[\sigma] \partial_\xi\theta_\ell 
  = -\frac{1}{D-2} \theta_k[\sigma] \theta_\ell[\sigma] 
    \left(\xi\partial_{\xi}q-q\right).
\label{eq:dktl}
\ee
The right-hand side of the cross-focusing equation becomes
\be
\begin{aligned}
&-\frac{1}{2}{\cal R} - \theta_{k}\theta_{\ell} + \omega^{2} 
    + {\cal D}\cdot\omega + \Lambda \\
 &\qquad = -\frac{1}{2}\xi^{2} {\cal R}[\sigma] 
    - \theta_{k}[\sigma] \theta_{\ell}[\sigma] q + \Lambda \\
&\qquad\;\;\;\;   + \left[\xi^{2D-2}-D\xi^{D+1}+(D-2)\xi^{D}+2\xi^{4}-\xi^{3}\right] 
    \left({\cal D}_{i}\log\theta_{k}[\sigma]\right)^{2} \\
&\qquad\;\;\;\;   - \left[2\xi^{2D-2}-D\xi^{D+1}+(D-2)\xi^{D}\right] \omega^{i}[\sigma] 
    {\cal D}_{i}\log\theta_{k}[\sigma] \\
 &\qquad\;\;\;\;  + \left(\xi^{3}-\xi^{D}\right)\Box\log\theta_{k}[\sigma] 
    + \xi^{D}{\cal D}\cdot\omega[\sigma] + \xi^{2D-2}\omega^{2}[\sigma],
\end{aligned}
\ee
so we have
\be
\begin{aligned}
  -\frac{1}{D-2}\xi\partial_{\xi}q + \frac{D-1}{D-2}q 
  &= -\frac{1}{2}\xi^{2} 
      \frac{{\cal R}[\sigma]}{\theta_{k}[\sigma]\theta_{\ell}[\sigma]} 
    + \frac{\Lambda}{\theta_{k}[\sigma]\theta_{\ell}[\sigma]} \\
  &\qquad + \left[\xi^{2D-2}-D\xi^{D+1}+(D-2)\xi^{D}+2\xi^{4}-\xi^{3}\right] 
    \frac{\left({\cal D}_{i}\log\theta_{k}[\sigma]\right)^{2}}
      {\theta_{k}[\sigma]\theta_{\ell}[\sigma]} \\
  &\qquad -\left[2\xi^{2D-2}-D\xi^{D+1}+(D-2)\xi^{D}\right] 
    \frac{\omega^{i}[\sigma]{\cal D}_{i}\log\theta_{k}[\sigma]}
      {\theta_{k}[\sigma]\theta_{\ell}[\sigma]} \\
  &\qquad + \left(\xi^{3}-\xi^{D}\right) 
      \frac{\Box\log\theta_{k}[\sigma]}{\theta_{k}[\sigma]\theta_{\ell}[\sigma]} 
    + \xi^{D}\frac{{\cal D}\cdot\omega[\sigma]}
      {\theta_{k}[\sigma]\theta_{\ell}[\sigma]} + \xi^{2D-2} 
      \frac{\omega^{2}[\sigma]}{\theta_{k}[\sigma]\theta_{\ell}[\sigma]}.
\end{aligned}
\label{eq:qeq}
\ee

We want to choose a gauge in which the zero $\xi_0(x^i)$ of $q$ (for $q$ solving 
\Eq{eq:qeq}) occurs at a uniform affine parameter, $\nu = \nu_0$ for all $x^i$ (i.e., $1/\xi_0(x^i ) = 1+\frac{\nu_0 \theta_k[\sigma]}{D-2}$, where the $x^i$-dependence in $\xi_0$ tracks the $x^i$-dependence in $\theta_k[\sigma]$),
thus making $Y_0$ a marginally-antitrapped surface, $Y_0 = Y_{\rm MA}$.  
That is, computing the zero $\xi_0(x^i)$ along each null generator, indexed by $x^i$, we need 
\be
{\cal D}_i \xi_0 = \left(\xi_0^2 - \xi_0\right){\cal D}_i \log \theta_k[\sigma]\label{eq:gaugeconditionxi}
\ee
for all $x^i$, as in \Eq{eq:Dxi}.
Let us first solve for $q$ in \Eq{eq:qeq} without making 
any a priori choice of the normalization of $k$ and then subsequently use 
gauge freedom to guarantee \Eq{eq:gaugeconditionxi} so that $\nu_0$ is independent of $x^i$.  Let us define the right-hand side of \Eq{eq:qeq} to be a function $f(\xi,x^i)$,
where the $x^{i}$-dependence enters only through the dependence of 
$\theta_{k}[\sigma]$, $\theta_{\ell}[\sigma]$, ${\cal R}[\sigma]$, and 
$\omega_a[\sigma]$ on their transverse position on $\sigma$.  The differential equation 
for $q$ can be written as
\be
  f   = -\frac{1}{D-2}\xi^{D}\partial_{\xi}\left(\frac{q}{\xi^{D-1}}\right),
\label{eq:feq}
\ee
which has solution
\be
  q(\xi) = -(D-2)\xi^{D-1} \int {\rm d}\xi\frac{f}{\xi^{D}},
\ee
where the integration constant is set by requiring $q=1$ at $\xi=1$.

Explicitly, defining $\psi_{i}[\sigma] = \omega_{i}[\sigma] - {\cal D}_{i} 
\log\theta_{k}[\sigma]$, we have
\be
\begin{aligned}
  q(\xi) &= \left( 1 - \lambda - \rho - \epsilon_{1} - \epsilon_{2} 
    - \epsilon_{3} - \epsilon_{4} - \epsilon_{5} \right) \xi^{D-1} \\
  &\qquad + \lambda + \rho\xi^{2} + \epsilon_{1}\xi^{3} + \epsilon_{2}\xi^{4} 
    + \epsilon_{3}\xi^{D} + \epsilon_{4}\xi^{D+1} + \epsilon_{5}\xi^{2D-2},
\label{eq:qeqfull}
\end{aligned}
\ee
where
\be
\begin{aligned}
  \lambda & =\frac{D-2}{D-1}\frac{\Lambda}{\theta_{k}[\sigma]\theta_{\ell}[\sigma]}\\
\rho & =-\frac{1}{2}\frac{D-2}{D-3}\frac{{\cal R}[\sigma]}{\theta_{k}[\sigma]\theta_{\ell}[\sigma]}\\
\epsilon_{1} & =\frac{D-2}{D-4}\frac{\Box\log\theta_{k}[\sigma]-({\cal D}_{i}\log\theta_{k}[\sigma])^{2}}{\theta_{k}[\sigma]\theta_{\ell}[\sigma]}\\
\epsilon_{2} & =2\frac{D-2}{D-5}\frac{({\cal D}_{i}\log\theta_{k}[\sigma])^{2}}{\theta_{k}[\sigma]\theta_{\ell}[\sigma]}\\
\epsilon_{3} & =-(D-2)\frac{{\cal D}\cdot\psi[\sigma]-(D-2)\psi^{i}[\sigma]{\cal D}_{i}\log\theta_{k}[\sigma]}{\theta_{k}[\sigma]\theta_{\ell}[\sigma]}\\
\epsilon_{4} & =-\frac{D(D-2)}{2}\frac{\psi^{i}[\sigma]{\cal D}_{i}\log\theta_{k}[\sigma]}{\theta_{k}[\sigma]\theta_{\ell}[\sigma]}\\
\epsilon_{5} & =-\frac{D-2}{D-1}\frac{\psi^{2}[\sigma]}{\theta_{k}[\sigma]\theta_{\ell}[\sigma]}.
\end{aligned}
\label{eq:constsfull}
\ee
Note that $\epsilon_{1,2,3,4,5}$ vanish for spherically-symmetric geometries in an appropriate gauge, 
while $\epsilon_{3,4,5}$ vanish if $\psi_{i}[\sigma]=0$.  In \Eq{eq:constsfull}, 
we have taken $D \geq 6$. For the special cases of $D = 3,4,5$, we 
can derive the analogues of \Eq{eq:qeqfull} and \Eq{eq:constsfull}, which 
we now compute.

\subsubsection{$D=3$}

For $D=3$, $\cal R$ vanishes, and the analogue of the right-hand side of \Eq{eq:qeq} is
\be
  f(\xi,x^i) = \frac{\Lambda}{\theta_k[\sigma]\theta_\ell [\sigma]} 
    + \xi^3 \frac{{\cal D}\cdot\omega[\sigma] - \omega^i[\sigma]{\cal D}_i 
      \log\theta_k [\sigma]}{\theta_k [\sigma] \theta_\ell[\sigma]} 
    + \xi^4 \frac{\omega^i [\sigma]{\cal D}_i \log\theta_k [\sigma] 
      + \omega^2[\sigma]}{\theta_k[\sigma] \theta_\ell[\sigma]},
\ee
so
\be
  q(\xi) = \left(1-\lambda-\chi-\tau\right)\xi^2 
    + \lambda + \chi\xi^3 + \tau\xi^4,
\label{eq:qD3}
\ee
where
\be
\begin{aligned}
  \lambda &= \frac{\Lambda}{2 \theta_k[\sigma] \theta_\ell[\sigma]} \\
  \chi &= -\frac{{\cal D}\cdot\omega[\sigma] - \omega^i [\sigma] {\cal D}_i 
    \log\theta_k[\sigma]}{\theta_k[\sigma] \theta_\ell[\sigma]} \\
  \tau &= -\frac{\omega^i[\sigma] {\cal D}_i \log\theta_k[\sigma] 
    + \omega^2[\sigma]}{2 \theta_k[\sigma] \theta_\ell[\sigma]}.
\label{eq:constsD3}
\end{aligned}
\ee

\subsubsection{$D=4$}

For $D=4$, the analogue of the right-hand side of \Eq{eq:qeq} is
\be
\begin{aligned}
  f(\xi,x^i) &= -\frac{1}{2} \xi^2 
      \frac{{\cal R}[\sigma]}{\theta_k[\sigma] \theta_\ell[\sigma]} 
    + \frac{\Lambda}{\theta_k[\sigma] \theta_\ell[\sigma]} \\
  &\qquad + \left(\xi^6-4\xi^5+4\xi^4-\xi^3\right) \frac{\left({\cal D}_i 
      \log\theta_k[\sigma]\right)^2}{\theta_k[\sigma] \theta_\ell[\sigma]} \\
  &\qquad - 2\left(\xi^6-2\xi^5+\xi^4\right) \frac{\omega^i[\sigma] 
    {\cal D}_i \log\theta_k[\sigma]}{\theta_k[\sigma] \theta_\ell[\sigma]} \\
  &\qquad + \left(\xi^3-\xi^4\right) 
      \frac{\Box \log\theta_k[\sigma]}{\theta_k[\sigma] \theta_\ell[\sigma]} 
    + \xi^4 \frac{{\cal D}\cdot\omega[\sigma]}
      {\theta_k[\sigma] \theta_\ell[\sigma]} 
    + \xi^6 \frac{\omega^2[\sigma]}{\theta_k[\sigma] \theta_\ell[\sigma]},
\end{aligned}
\ee
so
\be
  q(\xi) = (1-\lambda-\rho-\epsilon_{23}-\epsilon_4-\epsilon_5)\xi^3 
    + \lambda + \rho\xi^2 + \phi_1\xi^3\log\xi + \epsilon_{23}\xi^4 
    + \epsilon_4\xi^5 + \epsilon_5\xi^6,
\ee
where
\be
\begin{aligned}
  \lambda &= \frac{2\Lambda}{3\theta_k[\sigma]\theta_\ell[\sigma]} \\
  \rho &= -\frac{{\cal R}[\sigma]}{\theta_k[\sigma]\theta_\ell[\sigma]} \\
  \phi_1 &= -2\frac{\Box\log\theta_k[\sigma]-({\cal D}_i\log\theta_k[\sigma])^2}
    {\theta_k[\sigma]\theta_\ell[\sigma]} \\
  \epsilon_{23} &= -2\frac{4({\cal D}_i \log\theta_k[\sigma])^2 
    + {\cal D}\cdot\psi[\sigma] - 2\omega^i[\sigma]{\cal D}_i 
    \log\theta_k[\sigma]}{\theta_k[\sigma] \theta_\ell[\sigma]} \\
  \epsilon_4 &= -\frac{4\psi^i[\sigma]{\cal D}_i\log\theta_k[\sigma]}
    {\theta_k[\sigma]\theta_\ell[\sigma]} \\
  \epsilon_5 &= - \frac{2\psi^2[\sigma]}{3\theta_k[\sigma]\theta_\ell[\sigma]}.
\label{eq:consts4D}
\end{aligned}
\ee
Note that $\phi_1 = -\lim_{D\rightarrow 4}(D-4)\epsilon_1$, where $\epsilon_1$ 
is defined in \Eq{eq:constsfull}, and that $\epsilon_{23} = \epsilon_2 
+ \epsilon_3$ evaluated at $D=4$.

\subsubsection{$D=5$}

Finally, let us consider the special case of $D=5$.  The analogue of 
the right-hand side of \Eq{eq:qeq} is
\be
\begin{aligned}
  f(\xi,x^i) &= -\frac{1}{2} \xi^2 \frac{{\cal R}[\sigma]}
      {\theta_k[\sigma]\theta_\ell[\sigma]} 
    + \frac{\Lambda}{\theta_k[\sigma]\theta_\ell[\sigma]} \\
  &\qquad + \left(\xi^8-5\xi^6+3\xi^5+2\xi^4-\xi^3\right) 
    \frac{\left({\cal D}_i\log\theta_k[\sigma]\right)^2}
      {\theta_k[\sigma]\theta_\ell[\sigma]} \\
  &\qquad - \left(2\xi^8-5\xi^6+3\xi^5\right) 
    \frac{\omega^i[\sigma]{\cal D}_i\log\theta_k[\sigma]}
      {\theta_k[\sigma]\theta_\ell[\sigma]} \\
  &\qquad + \left(\xi^3-\xi^5\right) 
      \frac{\Box\log\theta_k[\sigma]}{\theta_k[\sigma]\theta_\ell[\sigma]} 
    + \xi^5 \frac{{\cal D}\cdot\omega[\sigma]}
      {\theta_k[\sigma]\theta_\ell[\sigma]} 
    + \xi^8 \frac{\omega^2[\sigma]}{\theta_k[\sigma]\theta_\ell[\sigma]},
\end{aligned}
\ee
so
\be
  q(\xi) = 
    \left(1-\lambda-\rho-\epsilon_1-\epsilon_3-\epsilon_4-\epsilon_5\right)\xi^4 
    + \lambda + \rho\xi^2 + \epsilon_1\xi^3 + \phi_2\xi^4\log\xi 
    + \epsilon_3\xi^5 + \epsilon_4\xi^6 + \epsilon_5\xi^8,
\ee
where
\be
\begin{aligned}
  \lambda &= \frac{3\Lambda}{4\theta_k [\sigma]\theta_\ell[\sigma]} \\
  \rho &= -\frac{3{\cal R}[\sigma]}{4\theta_k[\sigma]\theta_\ell[\sigma]} \\
  \epsilon_1 &= 3\frac{\Box\log\theta_k[\sigma] 
    - ({\cal D}_i\log\theta_k[\sigma])^2}{\theta_k[\sigma]\theta_\ell[\sigma]} \\
  \phi_2 &= -6\frac{({\cal D}_i\log\theta_k[\sigma])^2}
    {\theta_k[\sigma]\theta_\ell[\sigma]} \\
  \epsilon_3 &= -3\frac{{\cal D}\cdot\psi[\sigma] - 3\psi^i[\sigma] 
    {\cal D}_i\log\theta_k[\sigma]}{\theta_k[\sigma]\theta_\ell[\sigma]} \\
  \epsilon_4 &= -\frac{15\psi^i[\sigma]{\cal D}_i\log\theta_k[\sigma]}
    {2\theta_k[\sigma]\theta_\ell[\sigma]} \\
  \epsilon_5 &= -\frac{3\psi^2[\sigma]}{4\theta_k[\sigma]\theta_\ell[\sigma]}.
\label{eq:consts5D}
\end{aligned}
\ee
Note that $\phi_2 = -\lim_{D\rightarrow 5}(D-5)\epsilon_2$, where $\epsilon_{2}$ 
is given in \Eq{eq:constsfull}.

\subsection{Gauge fixing}
\label{subsec:gauge}

The surface $Y_0$ occurs at the first zero $\xi_0$ of $q$.  To require that 
the affine parameter $\nu=\nu_0$ at which this zero occurs to the same along every generator of $N_{-k}(\sigma)$, which would 
make $Y_0$ a bona~fide marginally antitrapped surface as 
required, we need \Eq{eq:gaugeconditionxi} to be satisfied.
Suppose we first compute $\xi_0$ as a function of $x^i$ and find that it does not satisfy \Eq{eq:gaugeconditionxi}, which would mean that $q$ does not vanish at constant affine parameter.
We can subsequently gauge transform the normalization of $k$ 
to enforce \Eq{eq:gaugeconditionxi}.  Let us define a rescaling of the vectors on $\sigma$ 
of the form
\be
\begin{aligned}
  k^{a} &\rightarrow e^{\Gamma} k^{a} \\
  \ell^{a} &\rightarrow e^{-\Gamma} \ell^{a}.
\end{aligned}
\ee
Then the affine parameter transforms as $\nu \rightarrow e^{-\Gamma}\nu$. Our $\xi$ parameter is invariant under this gauge transformation, $\xi \rightarrow \xi$.
However, the value of $\xi$ at which $q$ vanishes can change, since our various curvature quantities transform as
\be
\begin{aligned}
  \theta_k [\sigma] &\rightarrow e^{\Gamma} \theta_k [\sigma] \\
  \theta_{\ell}[\sigma] &\rightarrow e^{-\Gamma} \theta_{\ell}[\sigma] \\
  \omega_i[\sigma] &\rightarrow \omega_i[\sigma] + {\cal D}_i \Gamma \\
  {\cal R}[\sigma] &\rightarrow {\cal R}[\sigma] \\
  \psi_i[\sigma] &\rightarrow \psi_i[\sigma].
\end{aligned}
\label{eq:gauge}
\ee
Once we gauge fix so that \Eq{eq:gaugeconditionxi} is satisfied, we are guaranteed that 
$Y_0$, the surface on which $\theta_{\ell}=0$, is indeed marginally antitrapped.  

We can then construct an HRT surface by flowing along $N_{+\ell}(Y_0)$ 
as described in \Sec{subsec:buildHRT}.  For this construction to work, we need 
$\partial_\ell \theta_k < 0$ on $Y_0$.  The cross-focusing equation gives
\be
  \partial_\ell \theta_k 
  = -\frac{1}{2}{\cal R} - \theta_\ell \theta_k + \omega^2 
    - {\cal D}\cdot\omega + 8 \pi G\,T_{k\ell} + \Lambda 
  = \partial_k \theta_\ell - 2{\cal D}\cdot\omega.
\label{eq:dltk}
\ee
At $\xi_0$, we have $\partial_k \theta_\ell = -\frac{1}{D-2} 
\theta_k[\sigma] \theta_\ell[\sigma] \xi \partial_\xi q = \theta_k[\sigma] 
\theta_\ell[\sigma] f$ by \Eqs{eq:dktl}{eq:feq}.  Since $\xi_0$ by definition 
is the first zero of $q$ for $\xi > 1$ and $q(\xi=1) = 1$, we have 
$\partial_\xi q \leq 0$ at $\xi_0$, so it follows that $\partial_k \theta_\ell 
\leq 0$ at $\xi_0$.  By \Eq{eq:dltk}, the requirement that $\partial_\ell 
\theta_k < 0$ is a slightly different condition.  Provided this condition 
is satisfied, the area of the HRT surface is calculated from $\xi_0$:
\be
  A[X_{{\rm HRT}}] = \oint_\sigma \frac{\epsilon}{[\xi_0(x^i)]^{D-2}}.
\label{eq:areaHRTfinal}
\ee
where the integral is computed with the standard area $(D-2)$-form 
$\epsilon$ defined on $\sigma$ (so the area of $\sigma$ is just $A[\sigma] = \oint_\sigma \epsilon$).  

There are two conditions that must be satisfied for our construction of this 
HRT surface to work:
\vspace{-2mm}
\begin{enumerate}
\item There must exist a gauge transformation \eqref{eq:gauge} such that 
a solution $\xi_0$ of $q(\xi_0) = 0$ exists everywhere on $\sigma$ for 
$\xi_0(x^i)$ satisfying \Eq{eq:gaugeconditionxi}.
\label{zero}\vspace{-2mm}
\item We must have $\partial_\ell \theta_k [Y_0]<0$.
\label{minimar}
\end{enumerate}
Condition~\ref{zero} guarantees that we reach a $\theta_\ell = 0$ surface 
before $\theta_k$ diverges.  If, in a given gauge, $q(\xi) = 0$ cannot be 
satisfied along some null generator, it means that the geodesic in question 
hits a caustic before we reach a surface where $\theta_\ell$ vanishes. 
That is, one can show that condition~\ref{zero} guarantees that we have 
a one-to-one mapping along null generators from $\sigma$ to $Y_0$.%
\footnote{Specifically, suppose a geodesic from $\sigma$ undergoes a 
 nonlocal intersection with another member of the congruence between 
 $\sigma$ and $Y_0$; smoothness guarantees that the set of nonlocal 
 intersections in the congruence is bounded by caustics~\cite{Akers:2017nrr}. 
 Condition~\ref{zero} guarantees that such a caustic cannot occur to the 
 future of $Y_0$ along one of the null geodesics.  Moreover, if some part of $Y_0$ is 
 to the past of some nonlocal intersection but to the future of the caustic 
 (and hence to the future of other nonlocal intersections), then there must 
 be some geodesic with a nonlocal intersection on $Y_0$ itself.  This is 
 forbidden by definition of $Y_0$, since a nonlocal intersection on $Y_0$ 
 in the $k$ congruence would mean that both future-directed null vectors 
 have positive expansion, in contradiction with the requirement that one 
 of the future-directed null vectors have vanishing expansion on $Y_0$.
 \label{nonlocalfootnote}}
Moreover, the requirement in condition~\ref{zero} that $\xi_0$ satisfy \Eq{eq:gaugeconditionxi}
 is necessary to guarantee that the affine parameter corresponding to the zero of $q$ is independent of $x^i$, so that
 $Y_0$ is  a marginally-antitrapped surface as discussed in \Sec{subsec:buildHRT}. 
Finally, condition~\ref{minimar} is necessary to guarantee that $Y_0$ 
is a minimar surface in the sense of \Ref{Engelhardt:2018kcs}, so that 
we actually reach an HRT surface by flowing along $N_{+\ell}(Y_0)$. 
If one can freely solve the algebraic equation for $q(\xi) = 0$, then 
conditions~\ref{zero} and \ref{minimar} can all be checked using the data 
on $\sigma$.

\begin{figure}[t]
\begin{center}
\hspace{-5mm} \includegraphics[width=7cm]{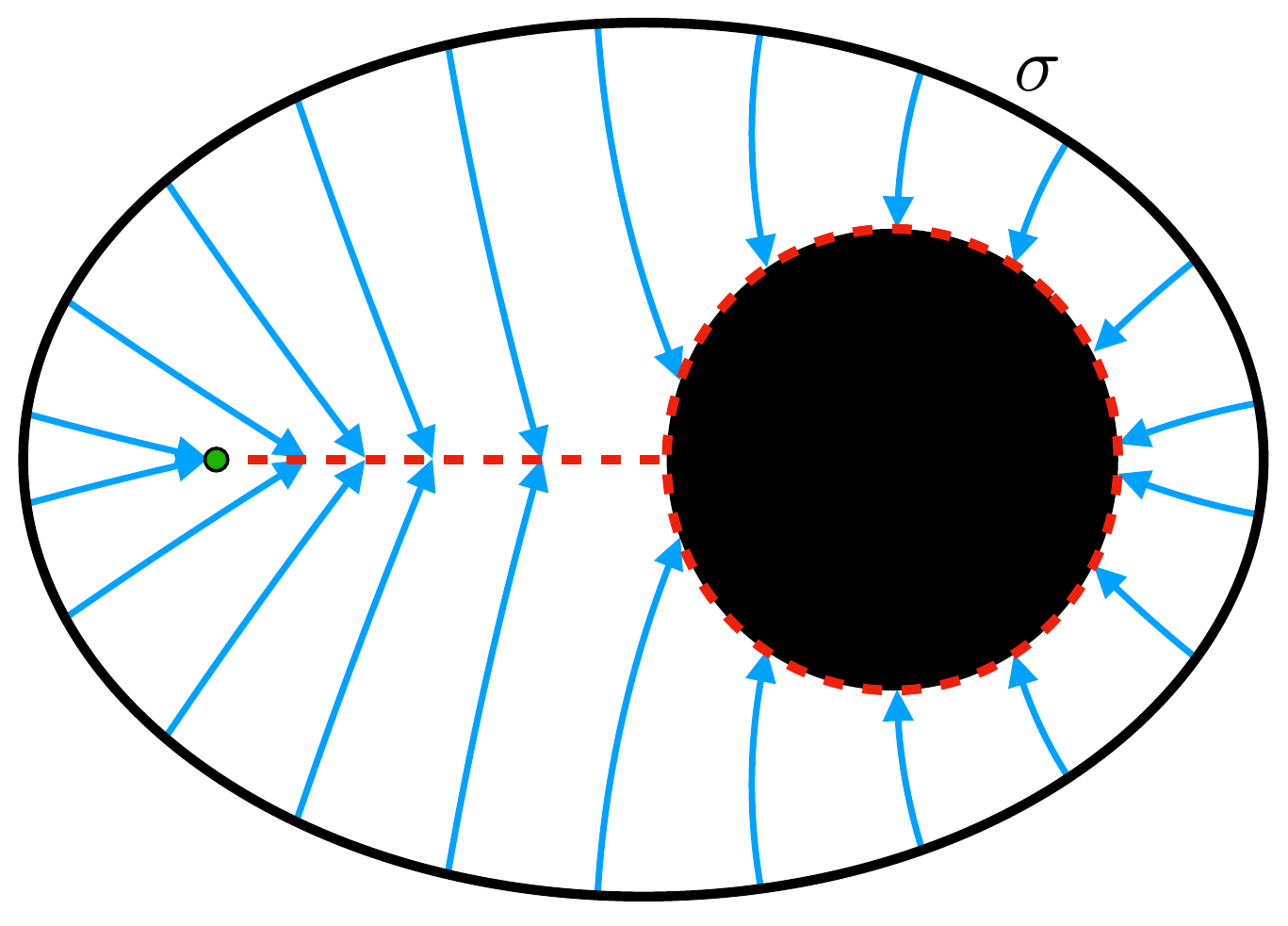}
\end{center}
\vspace{-5mm}
\caption{Illustration of a choice of a codimension-two surface $\sigma$ 
 (black line) that does not satisfy our veto condition~\ref{zero}.  The 
 light sheet in the $-k$ direction (blue arrows) unavoidably encounters 
 a caustic (green dot) along some generator before reaching the 
 marginally-antitrapped surface $Y_{\rm MA}$ (black circle).  Thus, 
 the surface (red dashed line) spanned by a slice of $N_{-k}(\sigma)$, 
 defined such that each generator either has $\theta_\ell = 0$ or 
 encounters a nonlocal intersection or caustic, has area larger than 
 the HRT surface.}
\label{fig:caustic}
\end{figure}

These conditions act as vetoes for surfaces $\sigma$: if $\sigma$ fails 
any of these conditions, our construction does not apply, and one must choose 
a different surface.  For a surface on which $N_{-k}(\sigma)$ unavoidably 
encounters caustics before reaching the $\theta_\ell = 0$ surface (see 
\Fig{fig:caustic}), we could imagine relaxing condition~\ref{zero} and 
instead merely find some maximal subset of the generators on $\sigma$ for 
which conditions~\ref{zero} and \ref{minimar} can be satisfied.  That is, 
if any geodesic cannot solve $q(\xi) = 0$, we can drop that geodesic, since 
it must reach a caustic before going through $Y_0$.  However, in this case, 
we do not have the guarantee discussed in footnote~\ref{nonlocalfootnote}, 
and we cannot rule out the possibility that some geodesics go through 
nonlocal intersections before encountering $Y_0$.  For such surfaces, 
our algorithm would therefore give an upper bound on the outer entropy (modulo the conjecture that the choice in \Eq{eq:zeros} is optimal).

More generally, one can compute the outer entropy for an arbitrary surface 
$\sigma$ failing condition~\ref{zero} without using our explicit algorithm, 
although such a computation would be challenging in practice.  For an arbitrary 
surface $\sigma$, consider the extension of $O_W(\sigma)$ to a spacetime 
${\cal M}$ for which the HRT surface interior to $\sigma$ is maximized. 
Then rather than using our explicit algorithm, one can define $Y_0$ to be 
the intersection of $N_{-\tilde \ell}(X_{\rm HRT})$ with $N_{-k}(\sigma)$, 
where $\tilde \ell$ is defined to be the ingoing null geodesic congruence 
orthogonal to $X_{\rm HRT}$, with zero shear.  If, as we have assumed, 
choosing $T_{ab}$ to vanish in $\overline{D}(\Sigma^+(\sigma))$ results 
in the optimal HRT surface, then the surface $Y_0$ exists in the spacetime, 
since the light sheet $N_{-\tilde \ell}(X_{\rm HRT})$ never ends and 
always has cross section with area equal to $A[X_{\rm HRT}]$.  If 
conditions~\ref{zero} and \ref{minimar} are satisfied, then there 
is a one-to-one correspondence between $Y_0$ and $\sigma$ induced 
by the geodesic congruence from $\sigma$ along the $-k$ direction. 
If condition~\ref{zero} fails, then one must relate $A[\sigma]$ and 
$A[X_{\rm HRT}]$ in the full spacetime by keeping track of which geodesics exit 
$N_{-k}(\sigma)$ between $\sigma$ and $Y_0$.  Even in this case, however, 
condition~\ref{minimar} is still needed, to guarantee that $\theta_k$ 
is positive on $Y_0$ so that $\sigma$ is a normal surface.

\section{Optimization}
\label{sec:optimization}

We now argue that our choices in \Eq{eq:zeros} indeed give the optimal HRT 
surface, so that the outer entropy is given simply by \Eq{eq:areaHRTfinal},
\be
  \So{\sigma} = \frac{1}{4G\hbar} \oint_\sigma \frac{\epsilon}{[\xi_0(x^i)]^{D-2}}.
\label{eq:equality}
\ee
This is one of the main results of this work: an algorithm for computing
the outer entropy (i.e., the area of the maximal HRT surface) for general 
codimension-two surfaces in general spacetimes.  We will give plausible 
physical arguments for why the choice \eqref{eq:zeros} should maximize the area of the 
HRT surface and hence conjecture that \Eq{eq:equality} holds, leaving a 
formal mathematical proof to future work.  Throughout, we assume the NEC, 
along with the version of the dominant energy condition that ignores the 
cosmological constant (dubbed the $\Lambda$DEC in \Ref{Nomura:2018aus}), 
which requires that $-T^a_{\;\;b}t^b$ be a future-directed, causal vector 
for all future-directed, causal $t^a$, so that the energy-momentum flow 
(excepting the cosmological constant) is causal in any reference frame. 
In particular, the $\Lambda$DEC implies that $T_{k\ell} \geq 0$, just as 
the NEC implies that $T_{kk}$ and $T_{\ell\ell}$ are nonnegative.

In the spherically-symmetric case, where the twist and shear vanish 
identically, the optimality of the choice $T_{kk} = T_{k\ell} = 0$, given 
the NEC and $\Lambda$DEC, was established in detail in \Ref{Nomura:2018aus}. 
Here, we simply mention that the reason for this can be inferred from 
the constraint equations~\eqref{eq:constraintk}: nonzero $T_{kk}$ would 
cause $\theta_k$ to grow more positive as we move toward the past along 
$N_{-k}(\sigma)$, and this would in turn increase $\nabla_k \theta_\ell$, 
which we want to engineer to be as negative as possible in order to reach the surface $X$ while 
incurring the least change in area from $\sigma$.

An essentially identical motivates us to take $T_{kk}$ and 
$T_{k\ell}$ to vanish in the general, nonspherical case.  Similarly, 
nonzero shear contributes to the Raychaudhuri equation in such a way as 
to accelerate the growth of $\theta_k$ along $N_{-k}(\sigma)$, counter 
to what we want for the construction, so we set $\varsigma_k$ to zero. 
As for the twist $\omega_a$, the ${\cal D} \cdot \omega$ term in the 
cross-focusing equation can contribute with either sign, but since its 
integral over any slice of $N_{-k}(\sigma)$ vanishes, it has no average 
effect on $\nabla_k \theta_\ell$ (though it can affect the global solution 
for $\xi_0$ due to its variation over $\sigma$).  On the other hand, the 
$\omega^2$ term has definite sign, making $\theta_\ell$ approach zero 
more slowly as we move along $N_{-k}(\sigma)$ and thereby decreasing 
$A[Y_0]$, which we do not want.  Once we have chosen $T_{ak} = \varsigma_k 
= 0$, the evolution of $\omega_a$ from its value on $\sigma$ is fixed by 
the DNS equation.  Therefore, to combat the deleterious 
effect of $\omega_a$, we could only imagine shutting off $\omega_a$ 
immediately to the past of $\sigma$ along $N_{-k}(\sigma)$ via a shock 
wave of nonzero $T_{ak}$ that cancels off $\omega_a[\sigma]$ precisely.%
\footnote{We cannot in general cancel off $\omega_a$ using nonzero 
 $\varsigma_k$ instead, since the $({\cal D}\cdot \varsigma_k)_a$ term 
 appearing in the DNS equation integrates to zero over 
 any codimension-two surface, while $\omega_a$ need not.}
However, as we will see below, this operation comes at a cost.

Let us define $v_a = -T_{a k}$, which the $\Lambda$DEC implies must be causal and 
future-directed, so $v^2 \leq 0$.  Since $g^{k\ell} = -1$, this implies 
$v_i v^i \leq 2 v_k v_{\ell}$.  Thus, 
\be
  g^{ij} T_{ik} T_{jk} \leq 2 T_{kk} T_{k\ell}.
\ee
In particular, a consequence of the $\Lambda$DEC is that setting $T_{kk} = 
T_{k\ell} = 0$ implies $T_{ak} = 0$ (and similarly, setting $T_{\ell\ell} = 0$ 
and $T_{k\ell} = 0$ implies $T_{a\ell} = 0$).%
\footnote{Moreover, the purely spatial components $T_{ij}$ can be similarly 
 bounded.  Define $t^a = \alpha \ell^a + \beta k^a + x^a$, where the unit vector $x^a$ 
 points in one of the transverse directions along $\sigma$ (so $x \cdot k 
 = x \cdot \ell = 0$).  The vector $t$ is timelike provided $2\alpha\beta > 1$. 
 Then defining $u_a = -T_{ab}t^b$, the $\Lambda$DEC implies that $u^2 \leq 0$.
 We find that if we have chosen $T_{kk} = T_{k\ell} = T_{\ell\ell} = 0$, which 
 means $T_{ak} = T_{a\ell} = 0$, then $u^2 \leq 0$ implies that $T_{ij} = 0$ 
 along all transverse directions $i,j$.  Hence, from the $\Lambda$DEC, we find 
 that choosing $T_{kk} = T_{k\ell} = T_{\ell\ell} = 0$ implies $T_{ab} = 0$.
 \label{Tabzero}}

Suppose that $8\pi G\, T_{ik}(\nu) = \delta(\nu) \omega_{i}[\sigma]$, 
corresponding to a shell of rotating matter.  By the DNS equation 
in \Eq{eq:constraintk}, the effect of this nonzero $T_{ik}$ is to zero 
out $\omega^2$ to the past of $\sigma$ along $N_{-k}(\sigma)$.  We 
saturate the $\Lambda$DEC by taking $8\pi G\, T_{kk} = \frac{x}{\sqrt{2}} 
\sqrt{\omega^2[\sigma]} \delta(\nu)$ and $8\pi G\, T_{k\ell} = 
\frac{1}{x\sqrt{2}} \sqrt{\omega^2[\sigma]} \delta(\nu)$ for some 
parameter $x$.  By the NEC, $x \geq 0$.  What does this shell of nonzero 
$T_{kk}$ and $T_{k\ell}$ do to $\theta_k$ and $\theta_\ell$?  It shifts 
them from their values on $\sigma$ to new values immediately to the past 
along $N_{-k}(\sigma)$.  That is, with $\nu = 0$ corresponding to $\sigma$, 
we have $\theta_k(\nu \rightarrow 0^-) = \theta_k[\sigma] + \Delta\theta_k$ 
and $\theta_\ell(\nu \rightarrow 0^-) = \theta_\ell[\sigma] + 
\Delta\theta_\ell$, where
\be
\begin{aligned}
  \Delta\theta_k &= 8\pi G\, 
    \int_{-\epsilon}^{\epsilon}{\rm d}\nu\, T_{kk}(\nu) \\
  \Delta\theta_\ell &= -8\pi G\, 
    \int_{-\epsilon}^{\epsilon}{\rm d}\nu\, T_{k\ell}(\nu).
\end{aligned}
\ee
Moving further along $N_{-k}(\sigma)$, to the past of these shifts, the 
solution proceeds in the same way as before, with $T_{kk}$ and $T_{k\ell}$ 
vanishing.  Hence, the cost of zeroing out $\omega^2$ is to shift $\theta_k$ 
and $\theta_\ell$.  Note that both shifts have signs that will decrease the area of $Y_0$, counter to our desired outcome.

A concrete example is illuminating.  Let us take an axisymmetric 
spacetime in $D = 3$, with $\sigma$ a circle centered on the origin, so that 
$\omega_a[\sigma]$ is a constant covector pointing in the angular direction and we choose a gauge in which $\theta_k$ and $\theta_\ell$ are constant over $\sigma$.  In this case, 
the $\chi$ term in \Eq{eq:constsD3} vanishes.  With our choice of nonzero 
$T_{ik}$ to cancel $\omega_i$, the $\tau$ term in \Eq{eq:constsD3} would also 
drop out, making $q = (1-\lambda')\xi^2 + \lambda'$, so the zero $\xi_0'$ is given by
\be
  \xi_0' = \left(1-\frac{1}{\lambda'}\right)^{-1/2},
\label{eq:chi0'}
\ee
where $\lambda'$ is $\lambda$ but with $\theta_k$ and $\theta_\ell$ shifted:
\be
  \lambda' = \frac{\Lambda}{2\left(\theta_k[\sigma]+\Delta\theta_k\right)
    \left(\theta_\ell[\sigma]+\Delta\theta_\ell\right)} 
  = \frac{\Lambda}{2\left[\theta_k[\sigma]\theta_\ell[\sigma]
    -\frac{\omega^2[\sigma]}{2}+\sqrt{\frac{\omega^2[\sigma]}{2}}
    \left(x\theta_\ell[\sigma]-\frac{1}{x}\theta_k[\sigma]\right)\right]}.
\ee
To minimize $\xi_0'$, we want $\lambda'$ to be maximized, which occurs when $x = 
\sqrt{-\theta_k[\sigma]/\theta_\ell[\sigma]}$, so
\be
  \lambda'=-\frac{\Lambda}{2\left(\sqrt{-\theta_\ell[\sigma]\theta_k[\sigma]}+\sqrt{\frac{\omega^{2}[\sigma]}{2}}\right)^2}.\label{eq:lambda'}
\ee

In contrast, if we instead take the construction of \Sec{sec:construction} 
with the choice of data given in \Eq{eq:zeros}, then we find the zero of 
$q(\xi) = (1-\lambda-\tau)\xi^2 + \lambda + \tau\xi^4$ (recalling that we 
are still taking $\chi = 0$ by axisymmetry) at
\be
  \xi_0 = \left[\frac{-(1-\lambda-\tau) \pm 
    \sqrt{\left(1-\lambda-\tau\right)^{2}-4\lambda\tau}}{2\tau}\right]^{1/2},
\label{eq:exD3chisol}
\ee
where $\lambda = \Lambda/2\theta_k[\sigma]\theta_\ell[\sigma]$ and 
$\tau = -\omega^2[\sigma]/2\theta_k[\sigma]\theta_\ell[\sigma]$ from 
\Eq{eq:constsD3}.  We choose the $-$~branch of the $\pm$ in \Eq{eq:exD3chisol} 
since we are interested in the smallest solution for $\xi\geq1$ (i.e., the 
first time $N_{-k}(\sigma)$ goes through a $\theta_\ell = 0$ surface).  Such 
a solution with $\xi_0 \geq 1$ exists if and only if
\be 
  \lambda \geq \left(1+\sqrt{\tau}\right)^2.
\label{eq:cond}
\ee

After some algebra, one can show using Eqs.~\eqref{eq:chi0'}, \eqref{eq:lambda'}, 
\eqref{eq:exD3chisol}, and \eqref{eq:cond}, along with the definitions of 
$\lambda$ and $\tau$, that $\xi_0$ is always strictly less than $\xi_0'$. 
Hence, the penalty in the shift of $\theta_k$ and $\theta_\ell$ outweighs 
any benefit from canceling off $\omega_i$, which means that our construction 
in \Sec{sec:construction} is better.  We conjecture that this example 
illustrates a general principle, namely, that the HRT surface interior to $\sigma$ 
is optimized by taking the background to have vanishing energy-momentum 
inside of $\sigma$.

Note that, given a minimar surface $Y_0$ as described in \Sec{subsec:buildHRT}, 
the HRT surface $X_{\rm HRT}$ that we eventually build by moving along 
$N_{+\ell}(Y_0)$ must, by definition, have area upper bounded by $Y_0$, 
as a consequence of the Raychaudhuri equation and the NEC.  Hence, the choice 
$T_{\ell\ell} = \varsigma_\ell = 0$ in \Eq{eq:zeros} was both necessary and 
sufficient to guarantee that $A[X_{\rm HRT}] = A[Y_0]$.  Moreover, while we 
constructed the HRT surface consistent with $O_W(\sigma)$ by moving first 
along the $k$ light sheet and then along the $\ell$ light sheet, we could 
have reversed the order, traversing $N_{+\ell}(\sigma)$ until we reached 
a surface $Z_0$ on which $\theta_k = 0$, choosing a gauge in which $Z_0$ 
is in fact a (marginally-trapped) minimar surface $Z_{\rm MT}$, and 
then traversing along $N_{-k}(Z_0)$ until we reach $X_{\rm HRT}$.  Under 
our assumption of \Eq{eq:zeros} that the HRT surface is optimized by 
choosing $T_{kk} = T_{k\ell} = T_{\ell\ell} = 0$ on $N_{-k}(\sigma)$ and 
$N_{+\ell}(Y_0)$, we found in footnote~\ref{Tabzero} that $T_{ab}$ must 
vanish identically on the past boundary of the inner wedge $I_W(\sigma) 
= \mathring D(\Sigma^+(\sigma))$ of $\sigma$.  Causality and conservation 
of energy-momentum then imply that $T_{ab}$ vanishes in the entirety 
of $I_W(\sigma)$.  Considering ${\cal M}$ to be an instantiation of a 
spacetime realizing the maximal HRT surface $X_{\rm HRT}$, which as noted 
in footnote~\ref{locationHRT} must be contained in $\overline{I}_W(\sigma)$, 
we can write the outgoing and ingoing orthogonal null congruences 
from $X_{\rm HRT}$ as $\tilde k$ and $\tilde \ell$, respectively, 
and define marginally-trapped and -antitrapped surfaces $Z_{\rm MT} 
= N_{+\tilde k}(X_{\rm HRT}) \cap N_{+\ell}(\sigma)$ and $Y_{\rm MA} 
= N_{-\tilde \ell}(X_{\rm HRT}) \cap N_{-k}(\sigma)$.  We can then choose 
a gauge in which $Z_{\rm MT} = Z_0$ or alternatively a (generally different) 
gauge in which $Y_{\rm MA} = Y_0$.  Under either gauge choice, we would 
manifestly construct the same maximal HRT surface, whether we applied our 
algorithm to the past or future boundary of $I_W(\sigma)$.  Hence, subject 
to the conclusions that we drew about the twist in the above section---that is, our 
assumptions about the optimality of requiring the vanishing of $T_{kk}$, 
$T_{k\ell}$ and $T_{\ell\ell}$---we conclude that the outer entropy is 
indeed given by our algorithm in \Sec{sec:construction}, so \Eq{eq:equality} 
holds for general spacetimes.

\section{Quasilocal Energy and Bekenstein-Hawking Entropy}
\label{sec:quasilocal}

As we have seen, our outer entropy $\So{\sigma}$ can be computed entirely in 
terms of curvature quantities ($\cal R$, $\theta_k$, $\theta_\ell$, $\omega_a$) 
defined on the codimension-two surface $\sigma$.  Hence, the outer entropy 
is a {\it quasilocal} quantity (cf.\ \Ref{Szabados:2004vb} and references 
therein); i.e., while not being a strictly locally-defined quantity, the domain 
on which it is computed is still finite.  Various other quasilocal quantities 
in general relativity can be defined.  Through a Gauss law argument for 
gravitational flux, such quasilocal quantities on codimension-two surfaces 
can be viewed as defining a notion of gravitational mass. 
In this section, we will find that the outer entropy itself admits an 
interpretation as such a quasilocal energy.  We will define the quasilocal 
energy in \Sec{subsec:quasilocaldef} and find that it exhibits several desirable 
features.  Subsequently, in \Sec{subsec:Hawking} we will explore the connections between the 
outer entropy and previously-defined quasilocal energies, including the Hawking 
mass~\cite{Hawking:1968qt,Hayward:1993ph}.

\subsection{Definition of a quasilocal energy}
\label{subsec:quasilocaldef}

Let us implicitly define a quasilocal energy $M$ by formally equating
$\So{\sigma}$ with the Bekenstein-Hawking entropy of a Schwarzschild 
black hole,%
\footnote{Throughout this section, we will work in $D \geq 4$ spacetime dimensions and will suppress $\hbar$.}
\be
  \So{\sigma} = \frac{\Omega_{D-2}}{4G} 
    \left[\frac{16\pi G M}{(D-2)\Omega_{D-2}}\right]^{\frac{D-2}{D-3}},
\label{eq:M}
\ee
recalling that the Schwarzschild radius of a $D$-dimensional black hole of ADM 
mass $M$ is $[16 \pi G M/(D-2)\Omega_{D-2}]^{1/(D-3)}$ and writing $\Omega_{D-2}$ for the area of the unit $(D-2)$-sphere.  That is, we are defining $M$ 
to be the mass of a Schwarzschild black hole of area equal to that of the largest HRT surface consistent with $O_W(\sigma)$. 
The expression in \Eq{eq:M} is defined precisely in analogy with 
the ``irreducible mass'' $m_{\rm irr}$ of a black hole with horizon area 
$A$~\cite{Christodoulou:1972kt,Szabados:2004vb},
\be
  A = \Omega_{D-2} 
    \left[\frac{16\pi G m_{\rm irr}}{(D-2)\Omega_{D-2}}\right]^{\frac{D-2}{D-3}}.
\label{eq:mirr}
\ee
Thus, we can view the mass $M$ defined in \Eq{eq:M}, corresponding to the 
outer entropy, as a definition of a new quasilocal energy in general relativity. 
In $D = 4$ dimensions, Eqs.~\eqref{eq:M} and \eqref{eq:mirr} reduce to 
$2GM = \sqrt{G\So{\sigma}/\pi}$ and $2Gm_{\rm irr} = \sqrt{A/4\pi}$.

Remarkably, our quasilocal energy $M$ is monotonic under inclusion.  This 
is a desirable property for an energy quantity in general relativity, but it 
is highly nontrivial from the perspective of the algorithm for computing 
$M$ (through $\So{\sigma}$) presented in \Sec{sec:construction}.  Rather, 
monotonicity under inclusion for $M$ arises as a consequence of the fact that 
$M$ defines an entropy.  By definition, $S^{\rm (outer)}$ grows monotonically 
under inclusion: for any new codimension-two surface $\sigma'$ containing 
$\sigma$ (i.e., for which $\sigma' \subset O_W(\sigma)$), we must have 
$\So{\sigma'} \geq \So{\sigma}$, since $O_W(\sigma') \subset O_W(\sigma)$ 
and so fewer degrees of freedom are being held fixed in $\So{\sigma'}$ than 
in $\So{\sigma}$ (that is, $\So{\sigma'}$ involves a maximization over a 
larger domain than $\So{\sigma}$).  Hence, assuming that our construction 
in \Sec{sec:construction} correctly computes the outer entropy, it follows 
that $M$ also grows monotonically under inclusion.

Our quasilocal energy $M$ also possesses other features one would want for a 
mass quantity in general relativity, including positivity, conservation, binding 
energy, and reduction to the irreducible mass for marginally-trapped surfaces, 
cf.\ \Ref{Szabados:2004vb}.  Since $\So{\sigma}$ is by definition nonnegative 
(and is manifestly so in \Eq{eq:equality}), $M$ is always real 
and nonnegative.  Further, since $M$ is quasilocal, as it is defined purely in 
terms of a codimension-two surface $\sigma$, it is by definition conserved if 
viewed as some energy integrated over a partial Cauchy slice passing through 
$\sigma$.  Moreover, since condition~\ref{zero} in \Sec{sec:construction} 
guarantees that points on $Y_0$ are mapped bijectively to points 
on $\sigma$ by the null congruence in the $k$ direction, it follows that 
$X_{\rm HRT}$ is topologically equivalent to $\sigma$.  Hence, for $\sigma$ 
consisting of two disjoint, closed components $\sigma_1$ and $\sigma_2$, 
the maximal HRT surface $X_{\rm HRT}(\sigma)$ is just the disjoint union of 
$X_{\rm HRT}(\sigma_1)$ and $X_{\rm HRT}(\sigma_2)$, so we have $\So{\sigma} 
= \So{\sigma_1} + \So{\sigma_2}$.  Since $M$ is a concave function of the 
black hole entropy, we 
have the strict inequality for the associated quasilocal energies,
\be
  M < M_1 + M_2.
\label{eq:subadditive}
\ee
Finally, for marginally-trapped surfaces, $\xi_0 \rightarrow 1$ and 
so the outer entropy computed in \Sec{subsec:solution} is simply 
$A[\sigma]/4G$~\cite{Engelhardt:2017aux}.  Hence, the quasilocal energy 
$M$ associated with the outer entropy in \Eq{eq:M} simply becomes the 
irreducible mass \eqref{eq:mirr}, i.e., $M = m_{\rm irr}$ for marginally-trapped 
surfaces.

\subsection{Hawking mass and beyond}
\label{subsec:Hawking}

It is instructive to compare $M$ to other proposed quasilocal energies in general 
relativity~\cite{Szabados:2004vb} and find limits in which they agree.  In $D = 4$ 
spacetime dimensions, the Hawking mass~\cite{Hawking:1968qt,Hayward:1993ph} is 
defined to be
\be
  m_{\rm Haw}[\sigma] = \frac{1}{8\pi G} \sqrt{\frac{A}{16\pi}} 
    \oint_\sigma \epsilon ({\cal R} + \theta_k \theta_\ell),
\label{eq:Hawking4D}
\ee
where $A$ denotes the area of $\sigma$ and as before the integral over $\sigma$ is computed with the standard area two-form 
$\epsilon$.  
We can 
infer the appropriate generalization of this expression to $D$ spacetime 
dimensions to be
\be
  m_{\rm Haw}[\sigma] = \frac{1}{8\pi (D-3)G} 
    \left(\frac{A}{\Omega_{D-2}}\right)^{\frac{1}{D-2}} \oint_\sigma \epsilon 
    \left( \frac{1}{2}{\cal R} + \frac{D-3}{D-2} \theta_k \theta_\ell\right),
\label{eq:HawkingD}
\ee
where $A$ is now the $(D-2)$-area of $\sigma$.  The Hawking mass is 
straightforward to compute for any given codimension-two surface, but, 
unlike our quasilocal energy derived from $S^{\rm (outer)}$, $m_{\rm Haw}$ 
is not in general positive or monotonic~\cite{Szabados:2004vb}.

In the spherically-symmetric limit, the four-dimensional Hawking 
mass \eqref{eq:Hawking4D} becomes the energy quantity of Misner, Sharp, 
and Hernandez~\cite{Misner:1964je,Hernandez:1966zia} and Cahill and 
McVittie~\cite{CahillMcVittie}:
\be
  m_{\rm MS}[\sigma] = \frac{1}{2G} r R^\phi_{\;\;\theta\phi\theta} 
  = \frac{1}{8G} r^3 R_{abcd} \epsilon^{ab} \epsilon^{cd} 
  = \frac{r}{2G}(1-g^{rr}).
\label{eq:MS4D}
\ee
We can develop a natural generalization of \Eq{eq:MS4D} to $D$ spacetime 
dimensions, writing
\be
  m_{\rm MS}[\sigma] = \frac{\Omega_{D-2}r^{D-1}}{32\pi (D-3)! G} R_{abcd} 
    \epsilon^{ab e_1 \cdots e_{D-4}} \epsilon^{cd}_{\;\;\;\;e_1 \cdots e_{D-4}} 
  = \frac{(D-2)\Omega_{D-2}r^{D-3}}{16\pi G} (1-g^{rr}).
\label{eq:MSD}
\ee
We indeed find that our $D$-dimensional generalization of the Hawking mass 
in \Eq{eq:HawkingD} reduces to our $D$-dimensional generalization of the 
Misner-Sharp energy \eqref{eq:MSD} in the spherical limit.  One can verify, 
for example, that by plugging in the $D$-dimensional Schwarzschild metric 
for which $g^{rr} = 1-\frac{16\pi G m}{(D-2)\Omega_{D-2} r^{D-3}}$, \Eq{eq:MSD} 
yields simply the Schwarzschild mass parameter, $m_{\rm MS} = m$.

Let us compare the Hawking mass to our outer entropy in the spherically-symmetric 
case.  Suppose we have a $D$-dimensional, spherically-symmetric spacetime 
($D \geq 4$) filled with pressureless dust plus a cosmological constant, 
with mass $m(r)$ inside radius $r$, so that
\be
  -g_{tt}(r) = g^{rr}(r) = 1 - \frac{2\Lambda r^2}{(D-1)(D-2)} 
    - \frac{16\pi G\,m(r)}{(D-2)\Omega_{D-2}r^{D-3}}.
\ee
We can identify a radius $R(r)$ implicitly defined as the largest solution of
\be
  1 - \frac{2\Lambda R^2(r)}{(D-1)(D-2)} 
    - \frac{16\pi G\,m(r)}{(D-2)\Omega_{D-2}[R(r)]^{D-3}} = 0.
\label{eq:defnR}
\ee
That is, if we collapse all of the matter interior to $r$, $R(r)$ is the radius 
of the resulting (A)dS-Schwarzschild black hole.  Let us find the outer entropy 
for a codimension-two shell at fixed $r$.  From \Sec{subsec:solution} and 
\Ref{Nomura:2018aus}, $\xi_0$ is the solution of
\be
  q(\xi_0) = (1-\rho-\lambda)\xi_0^{D-1} + \rho \xi_0^2 + \lambda = 0.
\ee
Recalling the definitions of $\rho$ and $\lambda$ from Eqs.~\eqref{eq:constsfull}, 
\eqref{eq:consts4D}, and \eqref{eq:consts5D}, we find $\rho = 1/g^{rr}(r)$ 
and $\lambda = r^2/L^2 g^{rr}(r)$, where for convenience we have defined 
$L^2 = -(D-1)(D-2)/2\Lambda$ for $\Lambda < 0$.  The solution to $q(\xi_0) = 0$ 
is $\xi_0  = r/R(r)$, as one can verify by plugging in the definition of 
$R(r)$ in \Eq{eq:defnR} and rearranging using the definition of $g^{rr}$.  
Hence, the outer entropy in \Eq{eq:equality} for this 
surface is
\be
  \So{\sigma} = \frac{\Omega_{D-2}[R(r)]^{D-2}}{4G}.
\ee
Namely, the outer entropy for the surface at $r$ is simply the Bekenstein-Hawking 
entropy one would obtain if all the matter (excluding the cosmological constant) 
were collapsed into a black hole.  The quasilocal energy $M$, according to 
\Eq{eq:M}, is then just the mass of a Schwarzschild black hole, with zero 
cosmological constant and radius $R$:
\be
  M[\sigma] = \frac{(D-2)\Omega_{D-2}}{16\pi G} [R(r)]^{D-3} 
  = m(r)\left[1-\frac{2\Lambda R^2(r)}{(D-1)(D-2)}\right]^{-1}.
\ee
The generalized Hawking mass from \Eq{eq:HawkingD} (or equivalently, 
$D$-dimensional Misner-Sharp energy in \Eq{eq:MSD}) associated with $\sigma$ is
\be
  m_{\rm Haw}[\sigma] = m_{\rm MS}[\sigma] = m(r) + \rho_\Lambda V_{D-1} r^{D-1},
\ee
where $\rho_\Lambda = \Lambda/8\pi G$ is the vacuum energy density and 
$V_{D-1} = \Omega_{D-2}/(D-1)$ is the Euclidean volume of the unit $(D-1)$-sphere. 
Thus, in the $\Lambda \rightarrow 0$ limit, we have
\be
  M[\sigma] = m_{\rm Haw}[\sigma] = m_{\rm MS}[\sigma] = m(r).
\ee
Since our construction required $T_{ab} = 0$ interior to $\sigma$, this matching 
is a consequence of Birkhoff's theorem. (For nonzero $\Lambda$, our quasilocal 
energy $M$ takes the cosmological constant into account differently than the Hawking 
mass.)  Specifically, if we take $\sigma$ to be a surface of arbitrary geometry subject 
to the constraint that it be topologically equivalent to a single sphere, centered in a spherical, static, 
asymptotically-flat spacetime with $T_{ab} = 0$ in $O_W(\sigma)$, Birkhoff's 
theorem~\cite{Birkhoff,Jebsen,Deser:2004gi} then guarantees that our quasilocal energy $M$ matches the ADM mass~\cite{Arnowitt:1959ah}
(or, equivalently in this case, the Bondi~\cite{Bondi:1962px,Sachs:1962wk} 
or Komar~\cite{Komar:1958wp} mass).

Hayward~\cite{Hayward:1993ph} introduced a modification of the Hawking mass that 
has the virtue of vanishing in flat spacetime (while the Hawking mass can be 
negative, even in Minkowski space).  The Hayward energy, $m_{\rm Hay}$, in $D = 4$ 
is defined by simply adding $-\frac{1}{2}(\varsigma_k)_{ab} (\varsigma_\ell)^{ab} 
- 2\omega^2$ to the integrand for the Hawking mass in \Eq{eq:Hawking4D}. 
Generically, our quasilocal energy $M$ will not match the Hayward energy, since 
as we saw in \Sec{subsec:solution}, $\So{\sigma}$---and hence $M$---depends 
in a complicated manner on derivatives of $\omega_a$, $\theta_k$, etc.\ on 
$\sigma$, in addition to $\omega_a$, $\theta_k$, etc.\ themselves.  However, 
$M$ and $m_{\rm Hay}$ share an important characteristic.  Like $m_{\rm Hay}$, 
$M$ will vanish in flat spacetime or pure (A)dS.  Specifically, starting 
with a surface in a nonvacuum spacetime that satisfies the conditions in 
\Sec{subsec:solution}, for which our algorithm computes the outer entropy, 
and taking the limit $T_{ab} \rightarrow 0$ in $O_W(\sigma)$, $\xi_0$ will diverge and so 
$S^{\rm (outer)}$ will go to zero.%
\footnote{This calculation was done explicitly for the spherical case in 
 \Ref{Nomura:2018aus} for Minkowski, AdS, and dS.  This conclusion follows in 
 general in the Minkowski case from the positive mass theorem~\cite{Schoen:1979rg,Witten:1981mf} 
 and in the (A)dS cases from its generalization to spacetimes that are not 
 asymptotically flat; see \Ref{Engelhardt:2017wgc} for an AdS/CFT perspective.}
On the other hand, while $m_{\rm Hay}$ is superadditive~\cite{Hayward:1993ph}---for 
$\sigma$ being the disjoint union of closed surfaces $\sigma_1$ and 
$\sigma_2$, one has $m_{\rm Hay}[\sigma] > m_{\rm Hay}[\sigma_1] + 
m_{\rm Hay}[\sigma_2]$---yielding a positive ``binding energy,'' the 
subadditive behavior of our quasilocal energy $M$ shown in \Eq{eq:subadditive} 
implies a negative binding energy $M - M_1 -M_2 < 0$, as one would physically 
expect.%
\footnote{However, unlike typical notions of gravitational binding energy, 
 both this binding energy and that of \Ref{Hayward:1993ph} are independent 
 of distance for distantly-separated surfaces.}

Finally, Liu and Yao~\cite{Liu:2003bx} and Kijowski~\cite{Kijowski} 
have defined a quasilocal energy $m_{\rm KLY}$ in $D = 4$ spacetime dimensions 
that exhibits positivity.  We will not discuss this energy in detail, except 
to comment that it differs from our $M$ in that $m_{\rm KLY}$ requires an 
embedding of $\sigma$ into flat three-dimensional space and furthermore, 
unlike $M$, does not equal the irreducible mass for marginally-trapped 
surfaces~\cite{Szabados:2004vb}.

\section{BTZ geometry}
\label{sec:BTZ}

An illuminating example in which the computation of the outer entropy manifests 
aspects of nonspherical spacetime while still maintaining tractability is 
the BTZ black hole geometry~\cite{Banados:1992wn}.  The line element for 
the $(2+1)$-dimensional black hole is ${\rm d}s^2 = -N^2 (r){\rm d}t^2 + 
{\rm d}r^2/N^2 (r) + r^2 \left(N_\phi (r){\rm d}t+{\rm d}\phi\right)^2$, 
where $N^2(r) = -M + \frac{r^2}{L^2} + \frac{J^2}{4r^2}$, $N_\phi (r) = 
-\frac{J}{2r^2}$, and the cosmological constant $\Lambda = -1/L^{2}$.  The 
angular momentum $J$ satisfies $|J| \leq M L$ for physical black holes.

We will consider a spacetime that, near some surface $\sigma$ at constant $r$, 
has a metric matching that of the BTZ black hole.  We will remain agnostic 
about the geometry of the spacetime inside or outside this surface.  Considering 
the geodesic congruences generated by the null vectors with initial 
tangents $k^a$ and $\ell^a$ orthogonal to $\sigma$, we can compute the null expansions, $\theta_{k}[\sigma] 
= -\theta_{\ell}[\sigma] = \frac{N(r)}{\sqrt{2}r}$, while the shears vanish 
identically for null congruences in $D = 3$, $\varsigma_k = \varsigma_\ell = 0$. 
Note that, if $r$ corresponds to a zero of $N(r)$, which occurs at the BTZ 
horizon
\be
  r_+ = L\sqrt{\frac{M}{2}\left[1+\sqrt{1-\left(\frac{J}{ML}\right)^{2}}\right]},
\ee
then expansions $\theta_k$ and $\theta_\ell$ vanish.  (The surface at $r = r_+$ 
can correspond to either the past or the future horizon.)

This spacetime exhibits a qualitative difference from the spherically-symmetric 
geometries considered by NR~\cite{Nomura:2018aus}: nonzero twist $\omega_a$. 
Computing the twist on $\sigma$ according to \Eq{eq:twistdef}, we find
\be
  \omega_a[\sigma] = \left(\frac{J^{2}}{4r^{3}}, 0, -\frac{J}{2r}\right).
\ee
Here, we have chosen the normalizations of $k^a$ and $\ell^a$ such that 
$\theta_k$, $\theta_\ell$, and $\omega_a$ are constant across $\sigma$. 
Note that this is not automatic; for example, we could replace $k^a \rightarrow 
e^{\Gamma(\phi)}k^a$ and $\ell^a\rightarrow e^{-\Gamma(\phi)}\ell^a$, for an 
arbitrary function $\Gamma(\phi)$, which would make the curvature quantities $\phi$-dependent.

To find the surface $Y_0$ where $\theta_\ell = 0$, we must find the first zero 
of $q(\xi)$ for which $\xi > 1$.  Here, $q(\xi)$ is given in \Eq{eq:qD3} for 
$D=3$.  Since $\omega_a$ and $\theta_k$ are constant across $\sigma$ under our 
chosen gauge, we have $\chi[\sigma] = 0$ in \Eq{eq:constsD3}, so $q(\xi)$ becomes
\be
  q(\xi) = \left(1-\lambda-\tau\right)\xi^2 + \lambda + \tau\xi^4,
\ee
where $\lambda$ and $\tau$ measure the cosmological constant and twist, 
respectively, as defined in \Eq{eq:constsD3}, which for the BTZ metric are 
$\lambda = \frac{r^2}{L^2 N^2(r)}$ and $\tau = \frac{J^2}{4r^2 N^2(r)}$.

The location of the zero in $q(\xi)$ is given by \Eq{eq:exD3chisol}.  For a subextremal
BTZ metric, the condition in \Eq{eq:cond} is satisfied for $r > r_+$, so a zero exists.  Plugging in the 
values of $\lambda$ and $\tau$ for our BTZ metric, we have $\xi_0^2 = 2M r^2 
\{[1-\sqrt{1-(J/ML)^2}]/J^{2}\}$.  As required by condition~\ref{zero} in 
\Sec{subsec:gauge}, $\xi_0$ satisfies \Eq{eq:gaugeconditionxi} everywhere on $\sigma$ in our 
gauge.  The area of $Y_0$, after some manipulation, is given by
\be
  A[Y_0] = \frac{2\pi r}{\xi_0} = 2\pi r_{+}.
\ee

We recall by the argument below \Eq{eq:dltk} that $\partial_k \theta_\ell[Y_0] 
\leq 0$.  Moreover, for our chosen congruence in this spacetime, ${\cal D} \cdot 
\omega = 0$, so by \Eq{eq:dltk} it follows that $\partial_\ell \theta_k[Y_0] 
\leq 0$.  More explicitly, the cross-focusing equation, along with our choices 
of initial data in \Eq{eq:zeros}, implies that, along $N_{+\ell}(Y_0)$, we have 
$\partial_{\ell}\theta_{k} = \omega^{2} + \Lambda$, which is constant by the 
DNS and Raychaudhuri equations along $N_{+\ell}(Y_0)$.  
At $Y_0$, we have, after some rearrangement,
\be
  \omega^2 [Y_0] + \Lambda 
  = \xi_0^4 \omega^2[\sigma] + \Lambda 
  = \frac{2M^2}{J^2} \left[1 - \left(\frac{J}{ML}\right)^2 
    - \sqrt{1-\left(\frac{J}{ML}\right)^2}\right] \leq 0,
\ee
with equality only in the extremal limit, $|J| \rightarrow ML$. If $|J| < ML$, 
$\theta_{k}$ is thus decreasing---at constant rate---along 
$N_{+\ell}(Y_0)$ and will eventually reach a surface where $\theta_k = 0$. 
Condition~\ref{minimar} in \Sec{subsec:gauge} is thus satisfied.  Since 
$\theta_k$ is constant over $Y_0$, the $\theta_k = 0$ slice of $N_{+\ell}[Y_0]$ 
occurs at constant affine parameter and hence corresponds to an HRT surface, as 
discussed in \Sec{subsec:buildHRT}.%
\footnote{In the extremal case, we have $\partial_\ell \theta_k[Y_0] = 0$, 
 so the minimar requirement of condition~\ref{minimar} does not hold and our 
 algorithm does not construct an HRT surface.}

Hence, the outer entropy associated with a surface $\sigma$, near which the 
geometry looks locally like subextremal BTZ, is just the Bekenstein-Hawking 
entropy of the corresponding BTZ black hole,
\be
  \So{\sigma} = \frac{2\pi r_+}{4 G\hbar}.
\ee
This was the result we expected. Indeed, in \Ref{AyonBeato:2004if}, an analogue 
of Birkhoff's theorem is proven for $(2+1)$-dimensional AdS gravity, where 
it is shown that all axisymmetric vacuum solutions of three-dimensional general 
relativity with negative cosmological constant and no timelike curves are either 
one of the BTZ geometries or the Coussaert-Henneaux~\cite{Coussaert:1994tu} 
spacetime.

\section{Discussion}
\label{sec:discussion}

In this paper, we have considered an interesting coarse-grained holographic 
quantity, the outer entropy, defined for general codimension-two surfaces. 
Using the characteristic initial data formalism describing the Einstein 
equations on light sheets, we have formulated  
an algorithm for constructing the optimal HRT surface consistent with the 
outer wedge, thereby calculating the outer entropy (\Sec{sec:construction}).  Motivated by examples, 
we have conjectured that the correct outer entropy is calculated by 
requiring that the interior of $\sigma$ have vanishing energy-momentum, other than the cosmological constant (\Sec{sec:optimization}). 
 Interestingly, we have found that the outer entropy 
offers a compelling definition of a quasilocal energy in general relativity. 
As discussed in \Sec{sec:quasilocal}, this quasilocal energy possesses several 
desirable features, including monotonicity under inclusion, positivity, 
binding energy, reduction to the irreducible mass for 
marginally-trapped surfaces, reduction to the Hawking and Misner-Sharp 
masses on spherical surfaces, and reduction to the BTZ mass for black holes 
in three dimensions.

This work leaves multiple promising directions for future research.  In our 
definition of the coarse-graining for the outer entropy, we have only held the 
spacetime degrees of freedom in the outer wedge $O_W(\sigma)$ fixed; that is, 
we have coarse-grained over all spacetime geometries outside of $O_W(\sigma)$, 
subject only to the constraints that they satisfy the Einstein equations, the NEC, 
and the $\Lambda$DEC.  However, it could be physically well motivated to somewhat 
fine-grain this requirement, depending on the matter sector of the theory.  In 
particular, if we add the further information that there are conserved charges 
in the theory, arising from some unbroken gauge field, then one could define 
a modified outer entropy in which we vary over all spacetimes satisfying the 
Einstein equation, energy conditions, and Maxwell's equations.  For example, 
if there is nonzero flux through $\sigma$, the question of whether and how 
quickly we can turn off $T_{k\ell}$ along $N_{-k}(\sigma)$---and whether 
doing so is to the benefit of our optimal HRT surface---hinges not only 
on the presence of the gauge field, but also on the spectrum of charged 
states in the matter sector.  If the theory contains an unbroken $U(1)$ 
gauge field but no charged matter (which violates the weak gravity 
conjecture~\cite{ArkaniHamed:2006dz,Cheung:2018cwt}), then $T_{k\ell}$ 
is unavoidably nonzero on $N_{-k}(\sigma)$ if there is flux through $\sigma$. 
Simultaneously solving the constraint equations and Maxwell's equations along 
the light sheet, one would then find that the area of the optimal HRT surface, 
and hence the outer entropy, would be lower.  This is to be expected, since 
adding information about the gauge field is in effect a fine-graining of the 
outer entropy definition, hence reducing the entropy.  It would be interesting 
to explore such modifications of the outer entropy in more detail.

In our construction of the HRT surface, we chose a gauge in which the surface 
$Y_0$ where $\theta_\ell$ vanished occurred at uniform affine parameter.  When 
the outer entropy was computed in the special case of marginally-trapped surfaces 
in \Ref{Engelhardt:2018kcs}, such a gauge choice was not made; instead, the fact 
that the congruence tangent $\ell$ did not in general equal the orthogonal null 
vector $\tilde \ell$ from the surface with $\theta_\ell = 0$ was accounted for 
by locating an alternative surface, on which $\theta_{\tilde \ell} = 0$, by 
relating $\theta_\ell$ and $\theta_{\tilde \ell}$ via a particular stability 
operator and then inverting it.  In our case, in which we are computing the outer 
entropy for more general surfaces, we could in principle construct---instead 
of solving the consistency equations for the gauge choice as described in 
\Sec{subsec:gauge}---the appropriate stability operator and solve the corresponding 
eigenvalue problem to relate $\tilde \ell$ and $\ell$ on $Y_0$.  However, the 
stability operator in \Ref{Engelhardt:2018kcs} is simplified by virtue of being 
anchored to a marginally (anti-)trapped surface.  The more general stability 
operator would be more mathematically complicated to invert; this difficulty 
should correspond to the challenge of solving the differential equations in 
\Sec{subsec:gauge}.  It could be worthwhile to further elucidate the connections 
between these two calculational methods.

By its definition as an entropy---or more specifically, as a maximization 
under a constraint---the outer entropy must satisfy a second law along the 
generalized holographic screens defined for non-marginally-trapped surfaces 
in \Ref{Nomura:2018aus}.  This is a manifestation of the growth of our 
quasilocal energy under inclusion, as discussed in \Sec{sec:quasilocal}, 
though demonstrating the entropy growth explicitly is highly nontrivial 
from the perspective of the algorithm given in \Sec{sec:construction}. 
In \Ref{Nomura:2018aus}, the rate of growth of the outer entropy along the 
generalized holographic screen was explicitly computed in the special case 
of spherical (but not necessarily marginally-trapped) surfaces; in addition 
to a second law, a Clausius relation was found, with the rate of change of 
the entropy being proportional to a certain flux in $T_{\mu\nu}$.  Investigating whether 
such a Clausius relation arises in the nonspherical case and more generally 
how to make the second law explicit from our algorithm could lead to 
a better understanding of the thermodynamic nature of the outer entropy 
for general surfaces.

As a new entry in the holographic dictionary, it would be interesting to 
investigate the CFT interpretation of the outer entropy for general surfaces. 
In \Ref{Engelhardt:2017aux}, it was shown that the outer entropy for 
marginally-trapped surfaces may be viewed as dual to a maximization of 
the boundary state under the action of certain ``simple operators.'' 
However, this interpretation relied crucially on the marginal-trappedness 
property of the surface under consideration.  From the perspective of the 
AdS/CFT dictionary, it would be good to understand how these definitions 
in the boundary theory are required to change for more general surfaces. 
We leave consideration of the boundary interpretation of our general outer 
entropy to future work.

\vspace{10mm}
 
\begin{center}
{\bf Acknowledgments}
\end{center}

\noindent
We thank Aron Wall for useful discussions and comments.  
This work was supported in part by the Department of Energy, Office of Science, 
Office of High Energy Physics under contract DE-AC02-05CH11231 and award 
DE-SC0019380, and by the National Science Foundation under grant PHY-1521446. 
Y.N. was also supported by MEXT KAKENHI Grant Number 15H05895, and G.N.R. 
by the Miller Institute for Basic Research in Science at the University 
of California, Berkeley.

\pagebreak

\bibliographystyle{utphys-modified}
\bibliography{Outer_entropy}

\end{document}